\documentclass{aa}  

\usepackage{graphicx}
\usepackage{txfonts}
\usepackage{graphicx}	
\usepackage{amsmath}	
\usepackage{amssymb}	
\usepackage{textgreek}
\usepackage{gensymb}
\usepackage{xcolor}
\usepackage[colorlinks=true, citecolor=blue]{hyperref}
\usepackage{etoolbox}

\begin{document} 

   \title{Revealing new high redshift quasar populations through Gaussian mixture model selection}

   \author{J. D. Wagenveld\inst{1,2}
         \and A. Saxena\inst{3}
         \and K. J. Duncan\inst{4,2}
         \and H. J. A. R\"{o}ttgering\inst{2}
         \and M. Zhang\inst{2}}

   \institute{Max-Planck Institut fur Radioastronomie, 
              Auf dem H\"{u}gel 69, 
              53121 Bonn, Germany
        \and
            Leiden Observatory, Leiden University, PO Box 9513, NL
            2300 RA Leiden, The Netherlands
        \and
            Department of Physics and Astronomy, University College London , 
            Gower Street, London, WC1E 6BT, UK
        \and
            SUPA, Institute for Astronomy, Royal Observatory, 
            Blackford Hill, Edinburgh, EH9 3HJ, UK
             }

\date{Received XX; accepted YY}

 
    \abstract{We present a novel method to identify candidate high redshift quasars (HzQs; ($z\gtrsim5.5$), which are unique probes of supermassive black hole growth in the early Universe, from large area optical/infrared photometric surveys. Using Gaussian Mixture Models to construct likelihoods and incorporate informed priors based on population statistics, our method uses a Bayesian framework to assign posterior probabilities that differentiate between HzQs and contaminating sources. We additionally include deep radio data to obtain informed priors. Using existing HzQ data in the literature, we set a posterior threshold that accepts ${\sim}90\%$ of known HzQs while rejecting $>99\%$ of contaminants such as dwarf stars or lower redshift galaxies. Running the probability selection on test samples of simulated HzQs and contaminants, we find that the efficacy of the probability method is higher than traditional colour cuts, decreasing the fraction of accepted contaminants by 86\% while retaining a similar fraction of HzQs. As a test, we apply our method to the Pan-STARRS Data Release 1 (PS1) source catalogue within the HETDEX Spring field area on the sky, covering $400$ sq. deg. and coinciding with deep radio data from the LOFAR Two-metre Sky Survey Data Release 1 (LoTSS DR1). From an initial sample of ${\sim} 5\times10^5$ sources in PS1, our selection shortlists 251 candidate HzQs, which are further reduced to 63 after visual inspection. Shallow spectroscopic follow-up of 13 high probability HzQs resulted in the confirmation of a previously undiscovered quasar at $z=5.66$ with photometric colours $i-z = 1.4$, lying outside the typically probed regions when selecting HzQs based on colours. This discovery demonstrates the efficacy of our probabilistic HzQ selection method in selecting more complete HzQ samples, which holds promise when employed on large existing and upcoming photometric data sets.}


\maketitle
%

\section{Introduction}

Studying large statistical samples of high-redshift quasars (HzQs) is essential for understanding the formation and evolution of super-massive black holes (SMBH) in the early Universe. The presence of Gunn-Peterson (GP) troughs \citep{gunn1965} in the spectra of HzQs at $z\sim6$, due to near-complete absorption of Ly$\alpha$ photons by the increasingly neutral intergalactic medium (IGM) along the line-of-sight, make them crucial probes of cosmic reionisation \citep[EoR;][]{fan2006, becker2015}. These GP troughs can in turn be used to photometrically identify large samples of HzQs, and the proliferation of wide area multi-band photometric surveys at optical wavelengths such as the Sloan Digital Sky Survey \citep[SDSS;][]{abazajian2003} and the Panoramic Survey Telescope and Rapid Response System surveys \citep[Pan-STARRS;][]{chambers2016} has enabled the discovery of statistically significant samples of bright quasars at high redshifts, with now over ${\sim}500$ confirmed HzQs at $z > 5$ \citep[see][for a compilation]{ross2020}.

For HzQs at $z\sim6$, towards the end of the EoR, the GP trough falls between the $i$- and $z$-band filters. Therefore, in the context of the SDSS and Pan-STARRS surveys (carried out using the $u, g, r, i, z$, and $y$ filters), quasars at $z\sim6$ may be identified through pin-pointing `$i$-dropout' sources, that show extreme $i-z$ colours. Searches for HzQs using photometric dropout techniques over large areas of the sky often employ linear cuts in magnitude and colours \citep[e.g.][]{banados2015,banados2016}. For example, a colour cut of $i-z > 1.5$ to $2.5$ is typically implemented in addition to magnitude cuts to ensure a balance between the selection of robustly detected HzQs and the exclusion of as many contaminating foreground sources as possible, which are often M, L and T-type brown dwarf stars in the Milky Way \citep{fan2001,willott2005,banados2016,jiang2016}. 

A radio detection can considerably aid in removing foreground contaminants such as dwarf stars that do not emit persistent radio continuum emission at the sensitivity of current observations \citep[e.g.][]{burningham2016}, as around $10\%$ of HzQs are seen to be `radio-loud' (radio loudness being defined as the ratio between rest frame radio and optical flux density) even out to high redshifts \citep[e.g.][]{banados2015}. However, overlapping deep radio data is often not available for the large sky areas covered by optical surveys from which candidate HzQs are selected. Deep radio continuum surveys of large sky areas, such as the Low-Frequency Array (LOFAR) Two Metre Sky Survey \citep[LoTSS;][]{shimwell2019}, can therefore potentially provide valuable additional information that could help improve HzQ selection and minimise the probability of contaminants. 

While selection based on optical and infrared colours from large area surveys has been highly successful in identifying some of the most distant HzQs currently known \citep[e.g.][]{fan2001, willott2010, banados2016, matsuoka2016, matsuoka2018, pipien2018, reed2019}, the use of linear cuts may lead to potential biases in the samples of HzQ candidates. A binary cut in colour and magnitude may inevitably lead to a loss of promising HzQ candidates. Additionally, \citet{mortlock2012} argued that linear cuts result in uniform grouping of high S/N candidates with more marginal ones that lie near the edges of the selection region, possibly making spectroscopic follow-up harder to prioritise. Finally, HzQs lying close to the limits of the redshift ranges probed by colour selections may be missed due to the GP trough not being fully sampled by the relevant broadband filters used for dropout selection. For example at redshifts of $z \sim 5.5$, the $i$-dropout selection may result in certain sources being missed, possibly presenting a gap in our understanding of SMBH evolution and/or the later stages of cosmic reionisation \citep{yang2017}. 

Additionally, binary selection criteria are often unable to properly account for the observational uncertainties in the observed properties for either individual sources or the population of sources being targeted. To overcome these limitations specifically in the case of identifying HzQs, a probabilistic selection as opposed to a binary selection may represent a better way to both obtain more complete samples of HzQs as well as assign higher probabilities for spectroscopic follow up to more promising candidates. \citet{bovy2011} introduced an implementation of Gaussian mixture modelling (GMM) that assumes and then deconvolves a model of the underlying population of sources from data, leading to a robust estimate of probability distribution of sources such as HzQs even from noisy measurements. Such an approach has been successfully employed to assign probabilities and better select low and intermediate redshift quasar candidates from SDSS data (e.g. \citealt{bovy2011}, but see also \citealt{bailer-jones2008, richards2009}).

Further complexities can be introduced in these models to improve the probability assignment, for example by also taking into account the respective prior probabilities of the different contaminants -- particularly dwarf stars in the Galaxy -- based on their spatial distribution and number densities on the sky. Such an approach was implemented by \citet{mortlock2012} for HzQs where prior information about populations that exhibit HzQ-like colours was used to assign a contamination (and as a result HzQ) probability and reduce the number of contaminants, leading to the discovery of a quasar at $z=7.1$ \citep{mortlock2011}. However, the initial selection of HzQ candidates in the probabilistic approach of \citet{mortlock2012} still relied on linear colour cuts, and could potentially suffer from the same incompleteness issues as faced by other colour-based HzQ selections.

Therefore, there remains room for improvement in probabilistic HzQ selection methods, by more accurately constraining the luminosity and sky distributions of possible contaminants to obtain more complete samples of HzQs. In this work we build upon the probabilistic approach of selecting HzQs based on posterior probability estimation using informed priors and likelihood estimation utilising GMMs. We also we make use of deep radio observations of the HETDEX spring field taken as part of the LoTSS first data release \citep[LoTSS DR1;][]{shimwell2017}, using radio detection as an additional prior to minimise foreground contamination. With the combination of multi-wavelength data and a probabilistic approach, we aim to develop a selection technique capable of uncovering more complete samples of HzQs from large area surveys, while minimising the number of contaminating sources present in these sample.

This paper is organised as follows. In Section \ref{sec:data} we describe the data sets that are used, and the here used HzQ selection method is described Section \ref{sec:hzq_selection}. In Section \ref{sec:candidate_sample} we apply our selection method to the data sets, obtaining probabilistically selected HzQ candidates. In Section \ref{sec:spec_obs} we present spectroscopic follow-up for a handful of high-priority HzQ candidates identified, and report the discovery of a previously undiscovered quasar at $z\sim5.7$. In Section \ref{sec:discussion}, we discuss the performance of our selection method, application to incoming large sky survey data sets and the possible implications of the discovery of P144+50. Finally, in Section \ref{sec:conclusions} we summarise the findings of this paper. 

Throughout this paper we assume a Planck 2015 cosmology \citep{planckCollaboration2015}, with $H_0 = 67.8\ \mathrm{km\  s^{-1}\ Mpc^{-1}}$, $\Omega_M = 0.308$, and $\Omega_{\Lambda} = 0.692$. All magnitudes are given in the AB system \citep{oke1983}, unless otherwise stated.

\section{Data}
\label{sec:data}

\subsection{Pan-STARRS}
\label{sec:ps1_data}
The primary data set used to identify HzQ candidates in this study is Pan-STARRS Data Release 1 (PS1). The PS1 survey covers $3\pi$ steradian of the sky, including the entire northern hemisphere \citep{chambers2016}, reaching $5\sigma$ depths of 23.3, 23.2, 23.1, 22.3, 21.3\,AB in the \textit{g,r,i,z} and {y} optical filters, respectively.

We first retrieve a sample of sources from the PS1 data archive\footnote{\url{https://panstarrs.stsci.edu}}, and although no colour cuts are made for the initial selection, a number of other criteria are applied to reduce the full PS1 sample down to the appropriate parameter space and more manageable numbers. Since we are primarily interested in HzQs, we require a non-detection in the $g$ and $r$ filters, while requiring a robust detection in the $i,z$, and $y$ filters. The non-detections are attributed to magnitudes fainter than the $5\sigma$ limiting magnitudes in the photometric filters published by the PS1 team, or values of $-999$ as this value is the magnitude assigned in case of a non-detection in a particular band. 

As a proof of concept, we also restrict our analysis to the sky area corresponding to the HETDEX Spring field, which is advantageously covered by deep radio data at 150 MHz from LoTSS DR1 (see Section \ref{sec:radio}). The selection criteria for obtaining an initial sample from the PS1 catalogue can thus be summarised as follows:
\begin{align*}
    &160\degree <\ \mathrm{R.A.}\ < 232\degree \\
    &42\degree <\ \mathrm{Dec.}\ < 62\degree \\
	&r_{P1} > 23.2 \ \mathrm{OR} \ r_{P1} = -999 \\
    &g_{P1} > 23.2 \ \mathrm{OR} \ g_{P1} = -999 \\
    &i_{P1} > 0,\ z_{P1} > 0, \ y_{P1} > 0 
\end{align*}
Furthermore, objects with flags from the PS1 processing pipeline indicating bad or low quality detections are excluded to remove sources that have poor photometric data, following Table 6 of \citet{banados2014}. From these criteria, a sample of ${\sim}5\times10^5$ sources with complete photometric data is retrieved. 

\subsection{Radio data}
\label{sec:radio}
The radio data used in this study is taken from LoTSS DR1 \citep{shimwell2017}, which is a low frequency radio continuum survey covering over 424 sq.deg of the northern hemisphere that reaches a median sensitivity of 71$\mu$Jy beam$^{-1}$. Consequently, radio sources that are considered are based on the $5\sigma$ LoTSS DR1 catalogues, have a flux density of at least 350 $\mu$Jy. Full details about the data reduction, processing, final images and source catalogues creation are presented in \citet{shimwell2019}, with robust optical cross identifications presented in \citet{williams2019} and accompanying photometric redshifts in \citet{duncan2019}. We additionally make use of early LOFAR `deep-fields' data in the Bo\"{o}tes field \citep{williams2018}, which coincides with the HETDEX Spring field sky area. 

Visual inspection of the radio and optical images of initial HzQ candidate samples drawn from PS1 demonstrated that dusty red galaxies at intermediate redshifts ($0 \lesssim z \lesssim 3$) represented an additional potential contaminant population that could also emit significant radio continuum emission. The Bo\"{o}tes deep field data from \citet{williams2018} therefore acts as a primary reference for those sources where the high-quality optical data enables robust photometric redshifts. 

\subsection{Ancillary data}

Several other data sets are utilised for training, testing, or validation of the GMM algorithms implemented in this work, or in accurate construction of priors. As a validation sample for HzQs, we use the sample of confirmed HzQs compiled by \citet{banados2016} containing all $z > 5.6$ quasars known as of March 2016. 

As a reference data set for dwarf stars in the Milky Way, we use a catalogue of brown dwarfs observed in PanSTARRS by \citet{best2017}. From the same work we use data on the mean absolute magnitudes of different dwarf types in PS1, which are also used for constructing their sky densities.

\section{High redshift quasar selection}
\label{sec:hzq_selection}

Having outlined the data sets that will be used to implement our new HzQ selection, in this section we describe the ingredients that go into the construction and implementation of our GMM based HzQ selection method.

Our HzQ selection method builds upon probabilistic selection of HzQs using a Bayesian framework presented by \citet{mortlock2012}, which does not rely on binary colour/magnitude cuts and incorporates additional prior knowledge about quasars and other contaminants to predict the likelihood of a source selected from a large area survey being a HzQ. For HzQs at $z \gtrsim 5.6$, therefore, a flexible algorithm can be constructed that can compute the probability for any given source based on its $i_{P1},z_{P1}$ and $y_{P1}$ magnitudes. 

We begin by first defining the posterior distributions for the classes of objects that are likely to occupy the photometric parameter spaces typically occupied by HzQs. We recall that the posterior probability of a source being part of any particular population can be calculated using Bayes' theorem:
\begin{equation}
	\mathrm{P}(C_k|X) = \frac{\mathrm{P}(C_k)\mathcal{L}(X|C_k)}{\sum_{i=1}^N \mathrm{P}(C_i)\mathrm{P}(X|C_i)}.
    \label{eq:bayes}
\end{equation}
where $P(C_k)$ is the prior probability of an object belonging to class $k$, and $\mathcal{L}(X|C_k)$ is the likelihood of the given source being part of class $k$, normalised over all $N$ possible classes and their associated probabilities.

In reality when considering the measurements of astronomical objects, additional factors related to both the distribution of sources on the sky as well as survey limitations must be accounted for when deriving probability estimates. More generally, considering these additional factors and the features $\mathbf{f}=\{f_1,f_2,...,f_n\}$ of a source that differentiates it from other sources in the data, the prior probability of sources belonging to class $k$ with parameters $\theta_k$ is calculated as:
\begin{equation}
	\mathrm{P}(C_k|\mathbf{f},\mathrm{det}) = \int \rho(\boldsymbol{\theta}_k)\mathrm{P}(\mathrm{det}|\boldsymbol{\theta}_k,C_k)\mathrm{d}\boldsymbol{\theta}_{k}\ .
\end{equation}
where $\rho(\boldsymbol{\theta}_k)$ is the sky density, and $\mathrm{P}(\mathrm{det}|\boldsymbol{\theta}_k,C_k)$ is the probability that the source is detected in the survey. 

For sources detected in a flux limited survey, the parameters $\theta_k$ most relevant to the probability are the magnitudes of the source classes in different filters, described by $\mathbf{m}_k$. In this case the features, $\mathbf{f}$, would describe the observed magnitudes of a given source in different filters, $\mathbf{\hat{m}}$. Therefore, to calculate the prior we marginalise over the relevant magnitude space. The prior is then combined with the likelihood in the `weighted evidence' term, describing the evidence that the source in question belongs to a given class:
\begin{equation}
	W_k(\mathbf{\hat{m}},\mathrm{det}) = \int \rho(\mathbf{m}_k)\mathrm{P}(\mathrm{det}|\mathbf{m}_k,C_k)\mathcal{L}(\mathbf{\hat{m}}|\mathbf{m}_k,C_k)\mathrm{d}\mathbf{m}_k\ ,
\end{equation}
where $\mathcal{L}(\mathbf{\hat{m}}|\mathbf{m}_k,C_k)$ is the likelihood of the features of a source belonging to an object of class $k$. With this, Eq.~\ref{eq:bayes} can be rewritten as
\begin{equation}
	\mathrm{P}(C_k|\mathbf{\hat{m}},\mathrm{det}) = \frac{W_k(\mathbf{\hat{m}},\mathrm{det})}{\sum_{i=1}^N W_i(\mathbf{\hat{m}},\mathrm{det})}.
\end{equation}
Having rewritten the equation to calculate the posterior probability of any given class of objects detected in a survey in terms of its apparent magnitude, below we describe the classes of sources that we consider in our search for complete samples of HzQs.

\subsection{Source classes}
\label{sec:source_classes}

Successful implementation of our HzQ selection method requires the proper identification of all classes of sources relevant that overlap with the HzQ parameter spaces. As a result, not every class of astrophysical source needs to be considered, which may also be considered as setting the prior of non-relevant source classes to zero. The relevant classes consist of the target HzQ population and a set of contaminating populations occupying the same feature space. Therefore, we identify three relevant populations: HzQs, dwarf stars within the Milky Way, and intermediate redshift dusty galaxies with red observed-frame optical colours. To model these populations we require data with PS1 magnitudes for each. 

As mentioned previously, we use the Galactic brown dwarf stars catalogue from \citet{best2017}, and the deep multi-wavelength galaxy catalogues in the Bo\"{o}tes field from \citet{williams2018} containing photometric measurements for intermediate redshift dusty galaxies. Both catalogues contain ${\sim}10^4$ sources, which is sufficient to model the colour space reliably without being biased by scatter in individual sources. 

The same, however, is not true for the catalogue containing ${\sim}$200 confirmed HzQs. Therefore, to model the distribution of the quasar population in the colour spaces probed, we simulate the rest-frame UV spectral energy distributions (SEDs) for a population of quasars using a distribution of power laws, $\alpha \sim \mathcal{N}(1.30, 0.38)$, following the distribution presented by \citet{cristiani2016}. These power law SEDs are then combined with emission lines using the SDSS quasar template from \citet{vandenberk2001}. 

To simulate then a population of high redshift quasars, each simulated quasar spectrum (continuum + emission lines) is redshifted. The redshift is drawn from the the redshift distribution following the co-moving luminosity functions as defined in \citet{mortlock2012}, Eq. (13), in the redshift range of $5.6 < z < 6.5$. A redshift dependent IGM absorption from \citet{madau1995} is then applied to simulated spectra, and the spectra are then convolved with the Pan-STARRS photometric filters 
(using prescriptions built into the \textsc{smpy} \textsc{python} package\footnote{\url{https://github.com/dunkenj/smpy}}). As we are only interested in obtaining a reasonable distribution in colour space, the results are not dependent on the absolute flux of the quasars. This method of generating quasar spectra results in a reliable distribution of HzQ colours and to maintain consistency with the number of contaminants available to model, we simulate a total of $10^4$ quasars in this manner. This method of simulating quasars rests on the assumption that both the \citet{vandenberk2001} template spectrum and power law distribution from \citet{cristiani2016} are valid to higher redshifts as well. While beyond the scope of this paper, more reliable samples of quasars could be generated using parametric SED modelling, which can account for intrinsic changes in quasar spectra as a function of luminosity and redshift \citep[e.g.][]{temple2021}  

\subsection{Likelihoods \& Gaussian Mixture Modeling}
\begin{figure*}
\centering
\includegraphics[width=\textwidth]{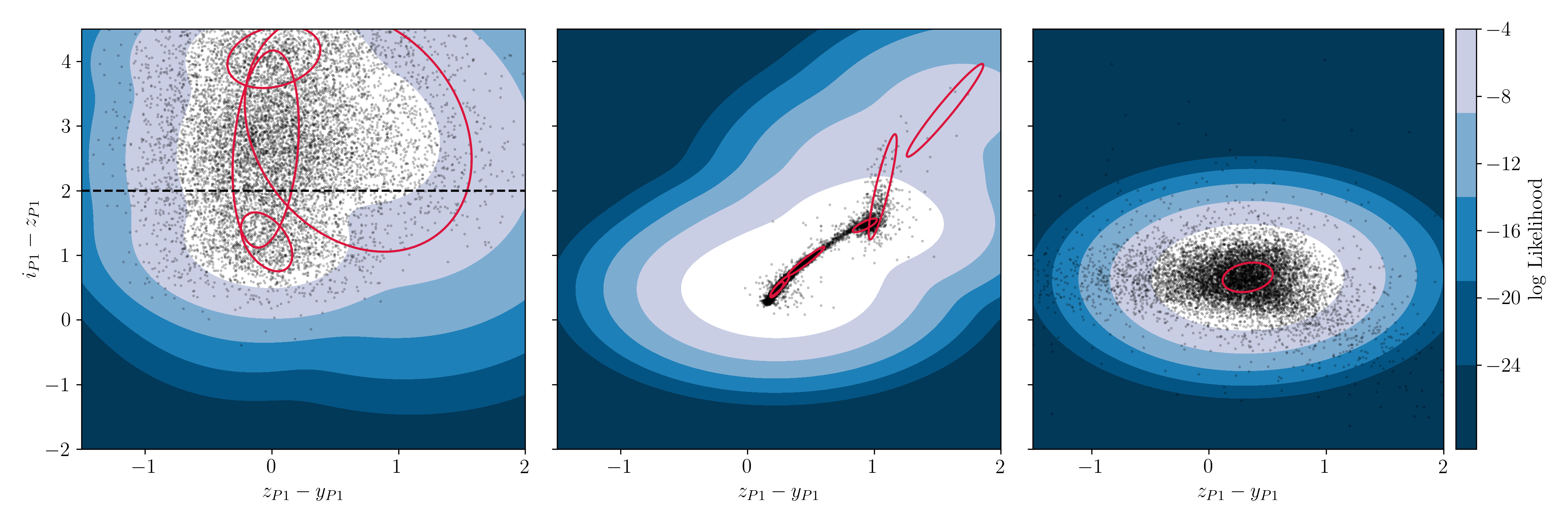}
  \caption{Log likelihood of quasars (left), dwarf stars (middle), and low redshift galaxies (right) modelled by a Gaussian mixture model in colour space. The red ellipses indicate the 3-sigma extents of the individual deconvolved Gaussians (4 components for quasars, 6 components for dwarf stars, 1 component for galaxies). The quasar likelihoods are assigned using photometry from simulated quasar spectra as highlighted in Section \ref{sec:source_classes}, whereas the photometry for dwarf stars and low redshift galaxies are taken from their respective reference catalogues, \citet{best2017} and \citet{williams2018}, respectively (black dots). The dashed line in the left panel marks the commonly defined colour cut at $i-z = 2.0$.}
  \label{fig:gmm}
\end{figure*}

In order to estimate the likelihood of a source belonging to a certain population that is considered in this study, we model each relevant population in the $i_{P1}-z_{P1}, z_{P1}-y_{P1}$ colour space. To do this, we use Gaussian mixture models (GMMs). A GMM assumes that the probability density of a population can be described by a finite number of weighted Gaussian functions \citep{reynolds2009}. Therefore, to obtain a probability density, $N$ Gaussian functions each with mean $\boldsymbol{\mu}_i$ and variance $\boldsymbol{\Sigma}_i$ are given a weight $w_i$, with the condition that the $N$ weights sum up to unity as follows:
\begin{align}
	& p(x) =  \sum_{i=1}^N w_i \mathcal{N}(\boldsymbol{\mu}_i,\boldsymbol{\Sigma}_i)\ , \\
	& 0\leq w_i \leq1, \ \ \sum_{i=1}^N w_i = 1\ .
\end{align}
The GMM is implemented in a machine learning algorithm, which optimises the parameters using Expectation-Maximisation \citep{dempster1977}. To estimate the number of components needed to model each population adequately, the Bayesian information criterion (BIC) is used \citep{wit2012}. This use of machine learning techniques to model various populations in colour space is a deviation from the method presented in \citet{mortlock2012}, and this is where the novelty of our method compared to traditional techniques relying on binary cuts in the colour space is highlighted best. The resulting likelihood of any given source belonging to a population follows directly from the GMM:
\begin{equation}
	\mathcal{L}(\mathbf{\hat{m}}|\mathbf{m}_{k},C_k) =  \sum_{i=1}^N w_i \mathcal{N}(\boldsymbol{\mu}_i,\boldsymbol{\Sigma}_i)\ .
\end{equation}
Furthermore, we use an extension of the classical GMM algorithm which implements extreme deconvolution, XDGMM \citep{holoien2017}. This implementation is particularly suited for noisy data, as it deconvolves the noisy distribution of the population in order to capture the underlying distribution more accurately. This method thus makes use of the uncertainties in the data, both for deconvolving the models and to assign likelihood to input data. As the error bars are folded into the covariance matrix of the GMMs, sources with larger uncertainties are assigned lower likelihood. The GMM algorithm is used to model the previously defined populations (quasars, dwarf stars, galaxies) in $i_{P1}-z_{P1}, z_{P1}-y_{P1}$ colour space. The log likelihoods (assuming a constant error in magnitude) of the resulting GMMs are shown in Figure \ref{fig:gmm}, along with the sources used to create the models for each population, in the left plot simulated quasars, and in the middle and right plots the sources from the reference catalogues \citep[][respectively]{best2017,williams2018}. The Gaussian components for each mixture model is also shown, with 4 components for HzQs, 6 components for dwarf stars, and one component for galaxies. The number of components that minimizes the BIC is chosen for each population separately.

\subsection{Detection Prior}
\label{sec:det_prior}

Many sources in our sample have faint magnitudes, extending all the way down to the PS1 detection limit. This makes obtaining accurate detection priors necessary to not only differentiate between real and fake sources, but also robustly characterise the various populations of sources considered, especially at the faintest magnitudes. 

Since the fraction of real sources detected as a function of source magnitude in PS1 \citep{metcalfe2013} is relatively well described by a sigmoid function, the detection priors we use for PS1 detected sources in this study are calculated as: 
\begin{equation}
	\mathrm{P}(\mathrm{det}|\mathbf{m}_k,C_k) = \left[1+\exp\left(4.84\cdot(m_{filt}-m_{filt,1/2}-0.4s)\right)\right]^{-1},
	\label{eq:detprior}
\end{equation}
where $m_{filt}$ is the magnitude of a source in one of the Pan-STARRS filters, $m_{filt,1/2}$ is the 50\% magnitude depth of said filter, and $s$ is a binary value that depends on the source type:
\begin{equation}
	s = 
	\begin{cases}
		0 &\mathrm{if\ star/point\ source}\\
		1 &\mathrm{if\ galaxy/extended\ source}
	\end{cases}
\end{equation} 
which is a relevant statistic for differentiating between point sources and extended sources. 

\subsection{Radio detection prior}
\label{sec:radio_prior}

Deep radio continuum data from LoTSS DR1 is used to complement the available optical data from Pan-STARRS, providing radio detections for a subset of the selected sources. To properly account for radio-detected sources, we modify the source classification based on the likelihood of radio detection, which we implement through the inclusion of an additional radio detection parameter, $f_{R,k}$, into the detection prior. Through this radio detection prior, if a radio counterpart in the LoTSS DR1 images of the input source is present, the radio detection is taken into consideration when computing the HzQ posterior probability.

For HzQs, roughly ${\sim}$10\% of the quasar population \citep[e.g.][]{hooper1995} is `radio-loud'. This relation seems to hold at higher redshifts, as \citet{banados2015} reported a radio-loud fraction of $\sim10\%$ for $z>5$ quasars at 1.4 GHz. Recent results from deep LOFAR survey data at lower redshifts have suggested that there is no dichotomy between radio-loud and radio quiet quasars, and that $30\%$ of quasars can detected by LOFAR surveys \citep{gurkan2019}. Similar fractions are found in LoTSS DR2 \citep[36\% at $>2\sigma$ significance;][]{gloudemans2021} at $z\gtrsim5.0$. A reasonable assumption therefore would be to set $f_{R,k} = 0.3$ for HzQs as the radio detection prior. 

For stars, including brown dwarf stars, the radio-loud fraction is very low, with \citet{kimball2009} finding about one in a million stars may be detected at radio wavelengths. However, low-frequency radio data combined with unparalleled sensitivity from LoTSS represents a new parameter space for the detection of radio signal from stars, as demonstrated by the recent discovery of polarised radio emission from a cold brown dwarf star \citep{vedantham2020}. Nevertheless, bright, non variable radio continuum emission sufficient to be detected in LoTSS imaging will be significantly rarer for brown dwarf contaminants than for luminous quasars or galaxies. Therefore, the probability of a radio detected source in our sample being a brown dwarf star is virtually zero, with $f_{R,k} = 10^{-6}$ for dwarf stars. 

For red, dusty galaxies at intermediate redshift, we find from the deep multi-wavelength catalogues based on deep LOFAR data in the Bo\"{o}tes field \citep{williams2018} that only a small fraction (${\sim}$1\%) of these galaxies has a radio detection. Therefore, we set $f_{R,k} = 10^{-2}$ for the galaxy population.

We note that these radio detection priors currently represent order of magnitude accuracy, and with deeper data collected over larger areas of the sky by current and future radio surveys, the radio detection priors can be improved upon to further enhance the probability assignment method for HzQs.

\subsection{Sky densities}

The sky densities of the source classes represent a significant prior, especially given the very rare nature of HzQs that makes any given source on the sky more likely to be a star or foreground galaxy. This prior can also differ depending on the apparent magnitude of the source and in this section we describe the calculation of the sky density priors for source populations considered in this study.

\subsubsection{M,L,T dwarfs}

Since the dwarf star contaminants are all within the Milky Way, the number density of dwarf stars at distance $d$ from the Earth and galactic latitude ($l$) and longitude ($b$) can be estimated assuming a galactic model \citep[e.g.][]{chen2001} as 

\begin{equation}
	n_i(d,l,b) = n_{0,i}\exp\left(-\frac{d\cos b \cos l}{h_R}\right)\exp\left(-\frac{|Z_{\odot}|+d\sin b}{h_Z}\right)\
    \label{eq:gal_dens_2}
\end{equation}
where $Z_{\odot}$ is the height of the Sun or Earth above the galactic plane, and $h_Z$ and $h_R$ are the characteristic height and distance scales for stars in the Milky Way, respectively \citep[see also][]{caballero2008}. The fiducial values of the various parameters used to calculate the sky densities of dwarf stars are given in Table \ref{table:dwarf_params}.

\begin{table}
	\centering
	\caption{Values of the various parameters used in Eq. (\ref{eq:gal_dens_2}) with errors as determined by \citet{chen2001}. The galactic latitude ($l$) and longitude ($b$) used here signify the centre of the HETDEX field.}
	\begin{tabular}{l c}
	\hline
	Parameter & Value \\
	\hline \hline
	$R_{\odot}$ & $8600\pm200$ pc \\ 
    $Z_{\odot}$ & $27\pm4$ pc \\ 
	$h_R$ & $2250\pm1000$ pc \\
	$h_Z$ & $330\pm3$ pc \\ 
	$l$ & 120 deg. \\
	$b$ & 65 deg. \\
	\hline
	\end{tabular}
    \label{table:dwarf_params}
\end{table}

Given the magnitude range specified, every dwarf type will have a slightly different heliocentric distance at which they will appear in the sample. To calculate this, the absolute PS1 magnitudes of each dwarf type are used from the \citet{best2017} catalogue. We calculate the sky density for each magnitude bin by integrating the spatial density over the cone covering the sky area. For a single stellar type, this results in
\begin{equation}
    \rho_i (\mathbf{m}_i) = \int_{d_1}^{d_2}n_i(D)D^2\mathrm{d}D,
\end{equation}
where $D$ is the distance in parsec. The total sky density of all contaminating dwarfs is thus calculated by the sum of the densities $\rho_i$ of all stellar dwarf types.

\subsubsection{Galaxies}

A significant fraction of sources in the PS1 data described in Section \ref{sec:data} are identified as faint red galaxies. As mentioned previously, the information for this population is primarily taken from the Bo\"{o}tes multi-wavelength photometric catalogue from \citet{williams2018}, also containing photometric redshifts that allows us to select such galaxies in the redshift range $0 \lesssim z \lesssim 3$. As this is a less well-defined astrophysical population compared to quasars and dwarf stars, we have no luminosity function to model their observed sky density. Instead, we use the apparent magnitudes of these galaxies in the PS1 $i$, $z$ and $y$ filters to model their distribution as a function of apparent magnitude in a given filter. We model the distribution with the Kernel Density Estimation (KDE) technique \citep{silverman2017}, where we use a bandwidth $h=0.1$ to get a smooth and continuous representation of the data. 

\begin{figure}
    \centering
    \includegraphics[width=\hsize]{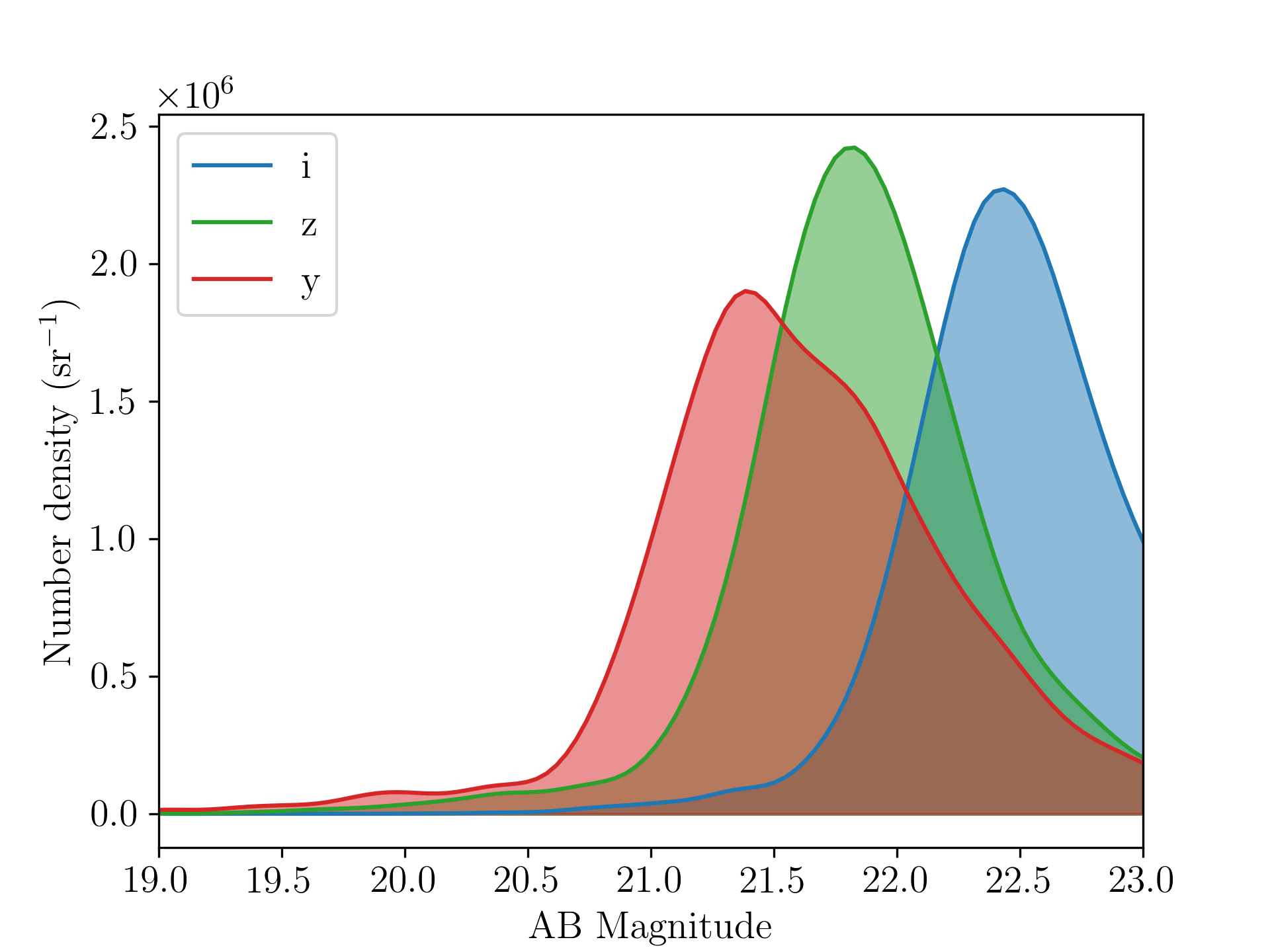}
    \caption{Expected number density of the galaxy populations as a function of apparent magnitude in PS1 $i$, $z$ and $y$ bands, extrapolated from the deep multi-wavelength galaxy catalogue in the Bo\"{o}tes field \citep{williams2018}.}
    \label{fig:gal_dens}
\end{figure}

This population of galaxies is made up of the population identified in the Bo\"{o}tes field, selected with the same criteria as the main PS1 sample (see Section \ref{sec:ps1_data}). The Bo\"{o}tes field covers an area of $S \simeq 11.6\,\mathrm{deg}^2$ on the sky, and we use this to convert the modelled distribution of galaxies to sky densities. Assuming that the galaxies are isotropically distributed, these sky densities are independent of the direction in which we observe, making the model valid for data in the HETDEX field as well. The resulting sky densities of galaxies detected in the PS1 $i$, $z$ and $y$ band data as a function of AB magnitude are shown in Figure \ref{fig:gal_dens}.   

\subsubsection{High redshift quasars}

Density functions of HzQs can be expressed in terms of the luminosity functions at high redshifts, which are poorly constrained compared to lower redshifts due to a lack of statistical samples \citep[e.g.][]{manti2017}. Using observations of quasars across redshifts, \citet{mortlock2012} derived a redshift and absolute magnitude dependent co-moving luminosity function for HzQs.

In order to calculate absolute magnitudes from the range of observed magnitudes in all relevant PS1 filters, $K$-corrections to the quasar spectra are calculated. The same method is used with which we simulated quasar magnitude in Section \ref{sec:source_classes}, by applying redshift dependent Lyman absorption from the intervening IGM based on redshift to the quasar SED templates from \citet{vandenberk2001}. We note that we do not account for the presence of ionised proximity zones around the HzQs. The Lyman absorbed and redshifted spectra are divided by the unaltered SED templates, and convolved with the relevant PS1 filters to obtain the $K$-corrections \citep[following][]{hogg2002}. 

Finally, integrating the redshift and magnitude dependent HzQ luminosity density, $\Phi_q(M,z)$, from \citet{mortlock2012} over the observed redshift cone yields the sky density of HzQs
\begin{equation}
    \rho_q (\mathbf{m}_q) = \int_{D_{\rm{co,1}}}^{D_{\rm{co,2}}}\, \Phi_q(M,z) \, D_{\rm{co}}^2 \, d D_{\rm{co}}
\end{equation}
where $D_{\rm{co}}$ is the co-moving distance in Mpc, integrated over the distances ($D_{\rm{co,1}}$, $D_{\rm{co,2}}$) corresponding to the redshift range probed.   

\subsection{Full posterior}

For the full photometric sample outlined in Section \ref{sec:data} we calculate the evidences for each class using the priors and likelihoods outlined above. The final quasar posterior probability is then constructed as
\begin{equation}
	\mathrm{P_q}(\mathbf{\hat{m}},\mathrm{det}) = \frac{W_q(\mathbf{\hat{m}},\mathrm{det})}{\sum_{i=1}^N W_i(\mathbf{\hat{m}},\mathrm{det})}
\end{equation}
where the weighted evidence is calculated based on the priors obtained using the $i$, $z$ and $y$ magnitudes and likelihoods in the $i-z$ and $z-y$ colour spaces for each source
\begin{align}
    \begin{split}
        	&W_k(\mathbf{\hat{m}},\mathrm{det}) = \\ 
        	&\int \rho_k(i,z,y)\mathrm{P}(\mathrm{det}|i,z,y,C_k)\mathcal{L}(\mathbf{\hat{m}}|i-z,z-y,C_k)\mathrm{d}i\mathrm{d}y\mathrm{d}z\ .
    \end{split}
\end{align}

As a result, every source with a measured $i$, $z$ and $y$ band magnitude in the PS1 catalogue can be robustly assigned a probability of being a HzQ. In the following section we apply our HzQ selection method to publicly available photometric data from PS1, in a bid to identify previously undiscovered HzQs at $z\gtrsim5.5$.

\section{Implementing the quasar selection algorithm}
\label{sec:candidate_sample}

Having defined all the necessary components for our HzQ probability assignment method, in this section we apply it to data taken from PS1 as described in Section \ref{sec:data}. When provided the $i_{P1},z_{P1}$ and $y_{P1}$ magnitudes of any source, our method described above should yield a posterior probability, $P_q$, that the source is a HzQ. 

\subsection{Candidate HzQ samples}
\begin{figure}
\centering
\includegraphics[width=\hsize]{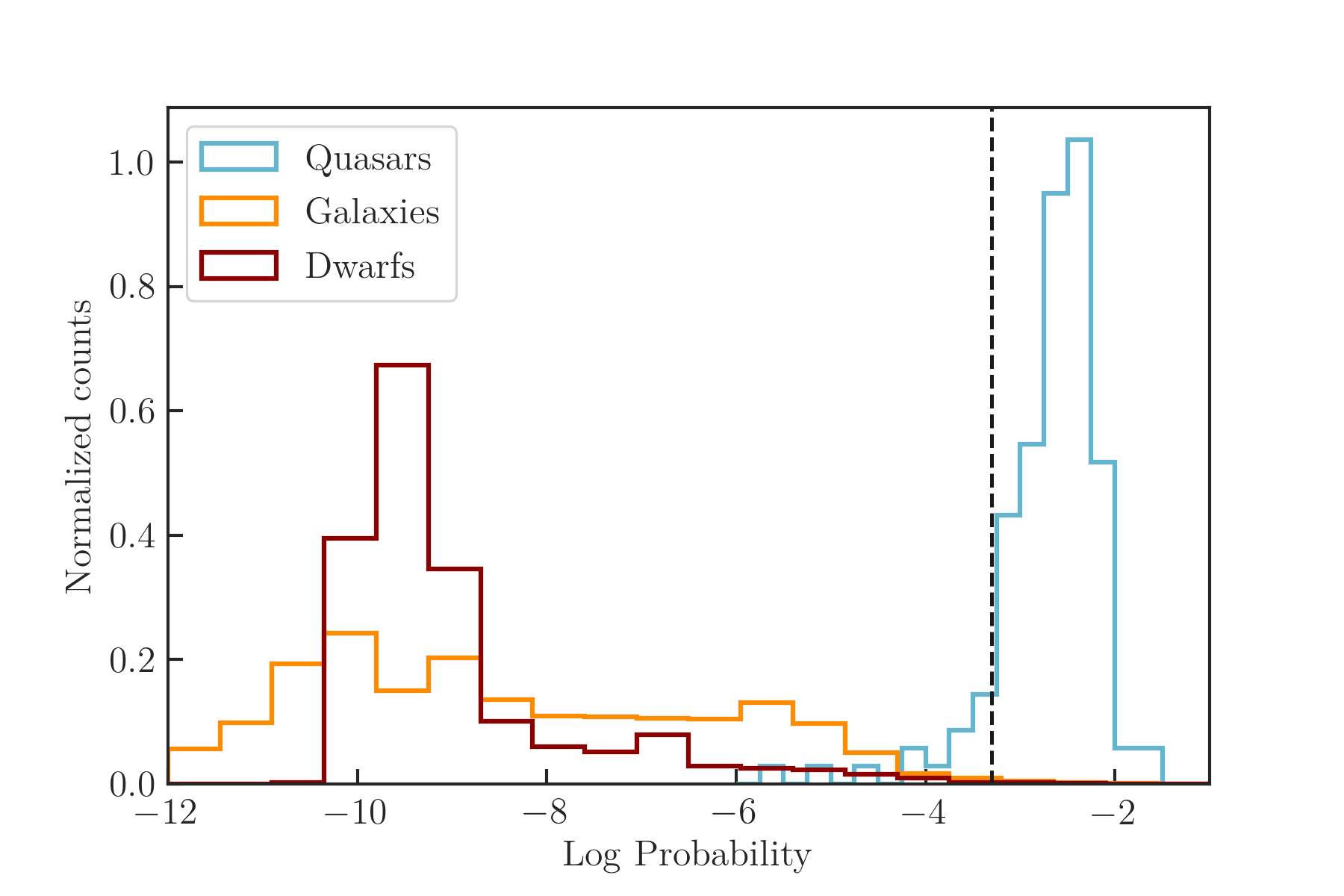}
  \caption{Posterior quasar probabilities of the different populations assigned by our HzQ selection method. The chosen posterior threshold of $P_q > 5\times10^{-4}$ is shown using the dashed line, which retains $\sim90\%$ of the known quasar population while rejecting $>99\%$ of foreground contaminant populations.}
     \label{fig:postprob}
\end{figure}

Due to the sky density priors giving a much greater weight to non-quasar likelihoods, we must define a posterior threshold that is capable of capturing the quasar population while largely rejecting other foreground contaminants. Using PS1 photometry of known quasars at $z>5.5$ from \citet{banados2016}, we find that a posterior threshold of $P_q > 5\times10^{-4}$ accepts ${\sim}90\%$ of the quasar population. The same threshold also rejects more than $99\%$ of dwarf stars and low redshift galaxies, as is shown in Figure \ref{fig:postprob}. 

We introduce an additional requirement ensuring a good quality detection in PS1 to remove spurious detections, saturated counts, and other instances resulting in bad photometry in the catalogue, by ensuring that the parameter \texttt{iQfPerfect} $> 0.85$. We exclude sources which do not adequately fit any of the modelled populations by removing sources which have low likelihood scores from all GMMs,
\begin{equation}
    \log \sum_k \mathcal{L}_k > -10\ .
\end{equation}
Having established the adequate posterior threshold that maximises the chances of identifying HzQs and minimises the incidence of foreground contaminants, we now proceed to run our novel HzQ selection method on the photometric data in the $i$, $z$ and $y$ bands queried from PS1. Our initial data set contained ${\sim}5\times10^5$ sources, out of which 508 sources were selected with probability above the set threshold. Finally, 263 sources satisfied the additional good quality detection requirement, of which 12 sources had an accompanying LOFAR detection.

\begin{figure}
    \centering
    \includegraphics[width=\hsize]{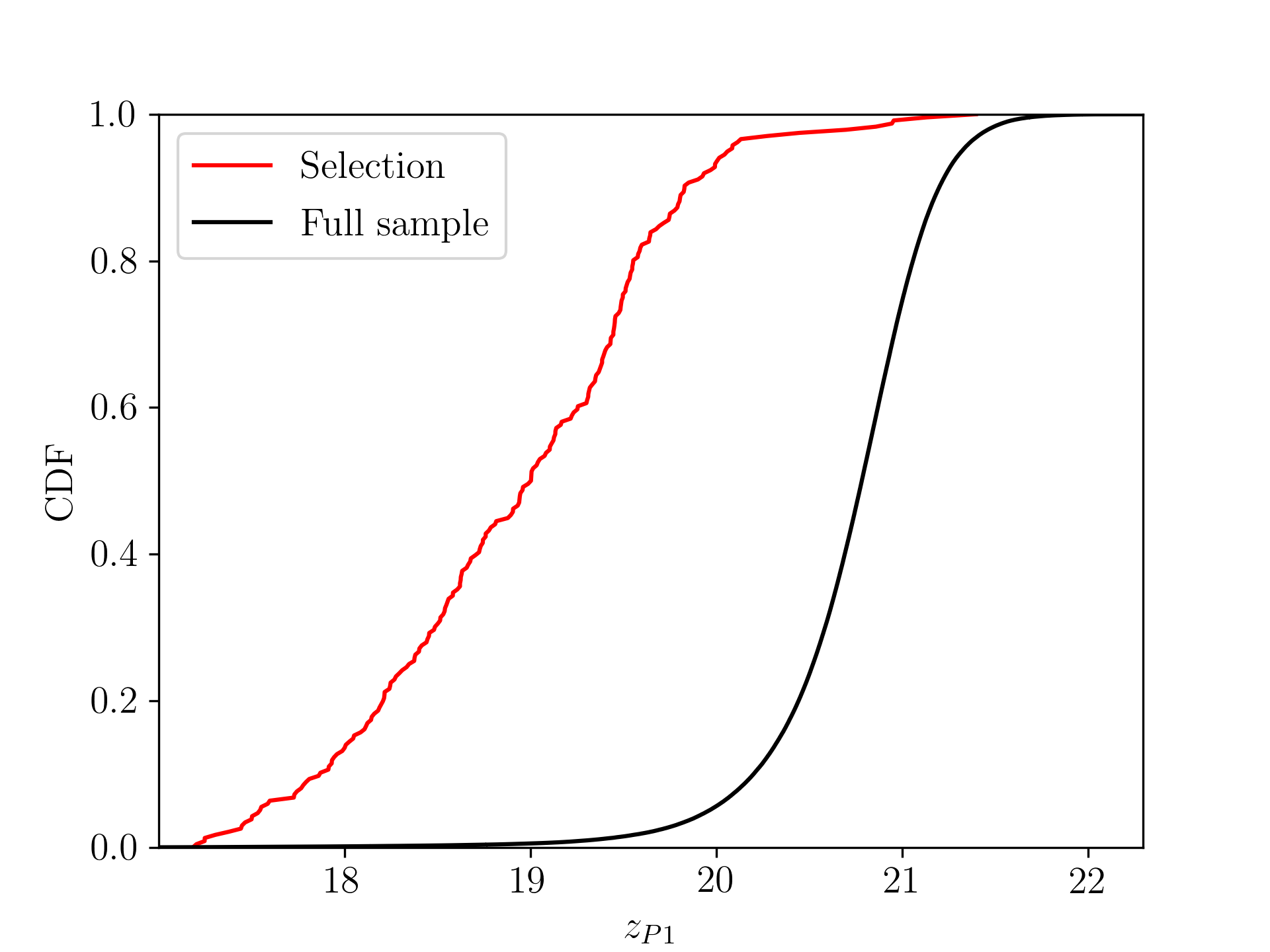}
    \caption{Cumulative distribution function (CDF) of $z_{P1}$ magnitudes of the sources classified as candidate HzQs (red line) compared with the full PS1 photometric sample (black line). Our algorithm preferentially assigns higher HzQ priorities to sources that are brighter and securely detected in the $z_{P1}$ band, ensuring that the Ly$\alpha$ break resulting in higher $i-z$ colours is securely constrained.}
    \label{fig:z_cdf}
\end{figure}

To investigate the selection function introduced by our algorithm to identify candidate HzQs, in Figure \ref{fig:z_cdf} we show the cumulative distribution function (CDF) of $z_{P1}$ magnitudes of sources lying above the probability threshold of being HzQs (red line), along with the CDF of the $z_{P1}$ magnitudes of all the sources that were passed through the algorithm (black line). Very clearly, our algorithm preferentially classifies objects with brighter $z_{P1}$ magnitudes job as candidate HzQs, placing a stronger emphasis on capturing the Ly$\alpha$ break which manifests itself as higher $i-z$ photometric colours. The brighter $z_{P1}$ magnitudes also ensure that sources classified as candidate HzQs are securely detected at redder wavelengths. The $z_{P1}$ mag distribution for sources with high HzQ probabilities peaks at $z_{P1} \approx 19$, with objects fainter than $z_{P1}>21$ very rarely selected, as it would be impossible to constrain the Ly$\alpha$ break in objects with the faintest $z_{P1}$ magnitudes. The comparison shown in Figure \ref{fig:z_cdf}, therefore, serves as a validation for our new HzQ selection algorithm.

The 263 candidate HzQs lying above the posterior threshold of $P_q > 5\times10^{-4}$ selected from our method represent a very small fraction ($\sim 0.05\%$) of the initial data set. Our strict threshold clearly results in a drastic reduction in the number of candidate HzQs, which can subsequently be visually inspected through which additional spurious, extended, or otherwise undesired sources can then be rejected. The main aim of visual inspection was to identify clearly spurious sources that may have been missed as such by the quality selection parameter (\texttt{iQfPerfect}). Examples include contamination by bright stellar spikes, cosmic ray residuals and grouped bright pixels. We additionally rejected candidates that showed extended morphologies, as HzQs are highly likely to appear as point sources in PS1 images. The visual inspection was carried out by JDW, AS and KJD, with mutual agreement being required in order to reject a candidate.

\begin{figure*}
\centering
\includegraphics[width=0.8\textwidth]{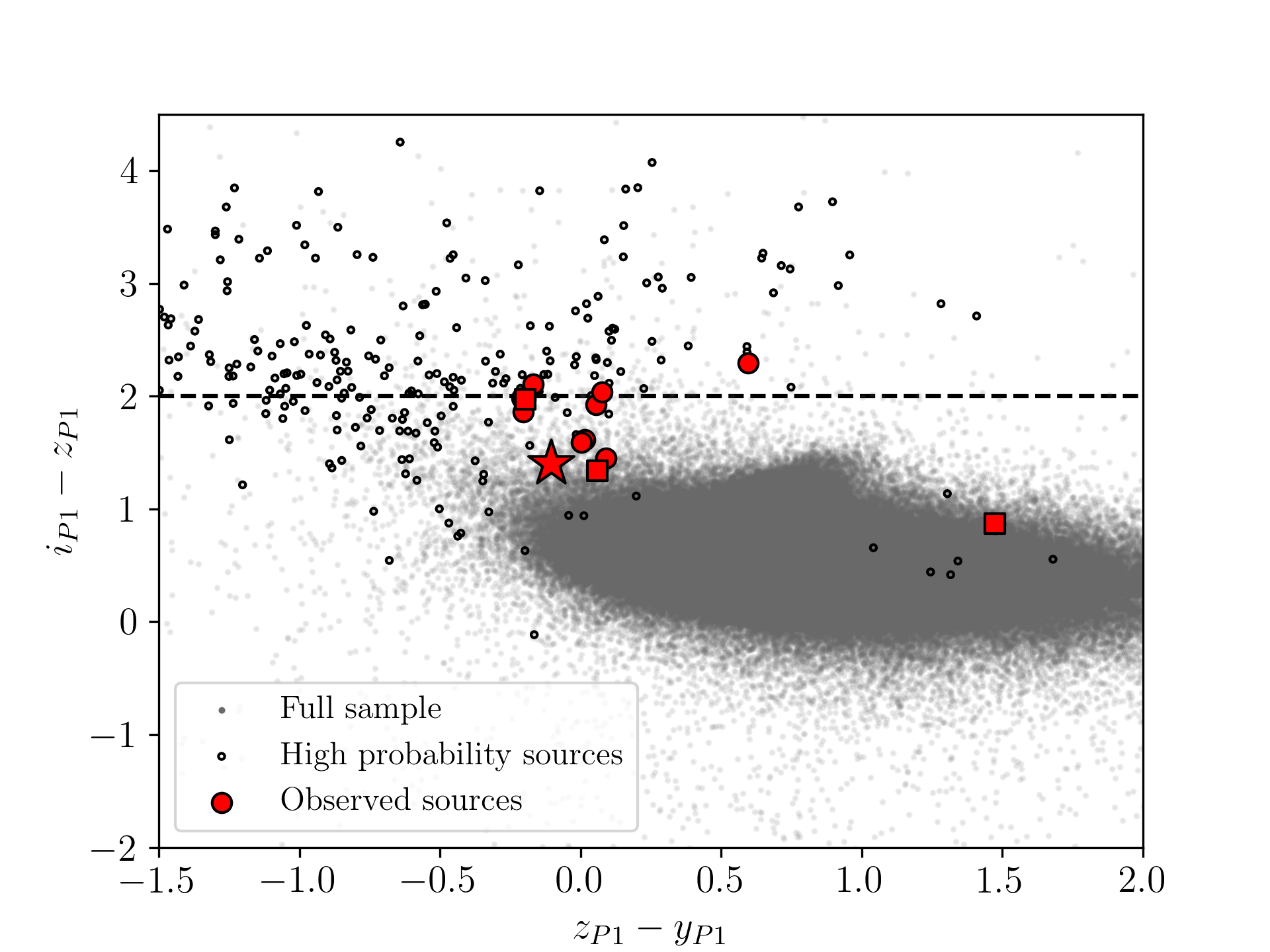}
  \caption{Distribution of sources in the $z_{\rm{P1}} - y_{\rm{P1}}$ vs $i_{\rm{P1}} - z_{\rm{P1}}$ diagram of the full PS1 sample (grey points), high probability HzQs (black circles) and the candidate HzQs selected for spectroscopic follow up (red circles) and those with a radio detection (red squares). The traditional colour cut at $i-z = 2$ is marked using the black dashed line, demonstrating that our method assigns a high HzQ probability to several sources that lie below this selection. We also mark P144+50 (red star) in the colour-colour plot, which was discovered using our method at a redshift of $z=5.66$ and with $i-z = 1.4$ lies below the standard photometric selection employed in other studies.}
    \label{fig:fullsample}
\end{figure*}

As a result, our conservative approach to visual inspection resulted in a large fraction of HzQ candidates being rejected, with 65 sources, 11 of which have a radio detection, remaining as good HzQ candidates suitable for spectroscopic follow up. The entire sample is summarised in Figure \ref{fig:fullsample}, where the discriminating power of the posterior calculation can be appreciated. Further details of the sources can be found in Table \ref{tab:candidates}. Noteworthy is a cluster of sources around $z-y = 1.5$, which represents a subset of the sources that have a detection in LoTSS. In total, 4417 sources in the full catalogue have a LOFAR counterpart, none of which would be selected if no radio counterpart was present. The addition of radio data has given higher significance to these sources, and shows that the method can robustly take into account the additional information provided by a radio detection.  

\subsection{Comparison with colour selection}

In this section we test the efficacy of our Bayesian HzQ selection method compared to the traditional colour and magnitude based selection.
We note that ${\sim}30\%$ of sources that were assigned high HzQ probabilities lie below the traditional $i-z > 2$ colour cut (Figure \ref{fig:fullsample}), and may potentially be missed by studies relying on binary colour/magnitude cuts for HzQ searches. To compare results, we apply a colour cut of $i-z > 2$ on the full PS1 sample, which selects 634 sources. This shows that even though sources below the colour cut can be above the probability threshold, many sources above the cut are also rejected. To investigate if these sources are rightfully rejected/accepted, we can run the algorithm on the data sets described in Section \ref{sec:source_classes}, and compare them with a $i-z > 2$ colour cut. 

For the HzQ sample, we simulate $10^4$ HzQs using the same method as described in Section \ref{sec:source_classes}. We also assign $z$-band magnitudes to the simulated quasars following a log-normal distribution based on the magnitude distribution of HzQs from \citet{ross2020}. From the $z$-band magnitude the $i$- and $y$-band magnitudes are automatically assigned based on the colours of the quasars. The algorithm is run on this sample and as for the PS1 sample, only sources above the probability threshold of $P_q = 5\times10^{-4}$ are accepted. Through this we find 7614 (76\%) HzQs above the probability threshold, while 7388 (74\%) are above the $i-z > 2$ colour cut. Much like in the sample described above, the probability cut rejects sources above the colour cut and vice versa, such that an important difference between the methods is in which parts of colour space are probed. 

As both Bayesian selection as well as colour selection methods return roughly the same number of HzQ candidates, we test the efficacy of selection by repeating the same experiment with the contaminant populations. For the brown dwarf population, we use the \citet{best2017} catalogue containing ${\sim}10^4$ sources, and using the above mentioned probability threshold selection, our algorithm shortlists 14 (0.15\%) known brown dwarfs as candidate HzQs. The colour selection, however, selects 68 (0.72\%) brown dwarfs as HzQ candidates.

Using the low/intermediate redshift galaxy catalogues from \citet{williams2018} containing ${\sim}10^4$ sources, our algorithm classifies only 1 (0.01\%) galaxy as a candidate HzQ, compared to 36 (0.38\%) galaxies being classified as HzQ candidates based on colour selection. Overall, we find that our method rejects a larger percentage of contaminants, while retaining a similar fraction of HzQs compared to a simple $i-z > 2$ colour selection, implying a higher overall efficacy. 

We note that in $i-z$ colour based selections, often an additional colour criterion of $z-y < 0.5$ is applied \citep[e.g.][]{banados2014} to further remove contaminants. Applying these $i-z > 2$ and $z-y < 0.5$ cuts on our simulated HzQs, brown dwarf and galaxy catalogues, we find that all brown dwarfs are eliminated and 28 galaxies are classified as candidate HzQs. However, these cuts only retain $58\%$ of HzQs from our simulated sample, clearly showing that although a large fractions of contaminants are eliminated by using colour cuts based on $i$, $z$ and $y$ band photometry, several HzQs may also be missed by such a selection. 

The possible addition of radio data further improves the efficacy of HzQ selection, and we consider a sample where all sources have counterparts in the radio. Using the assumed radio detection rates in Section \ref{sec:radio_prior}, we see almost all quasars with a radio detection are accepted (91\%). This fraction is purely considering the amount of sources that are above the probability threshold, which is in addition to the increase in probability for these quasars across the board. Effectively all dwarf stars are eliminated from the sample, while as before 0.01\% of galaxies are retained. This shows that the addition of radio data can be extremely valuable to identifying HzQs, and significantly increases the purity of the resulting candidate sample. 

\begin{figure*}
    \centering
    \includegraphics[width=\textwidth]{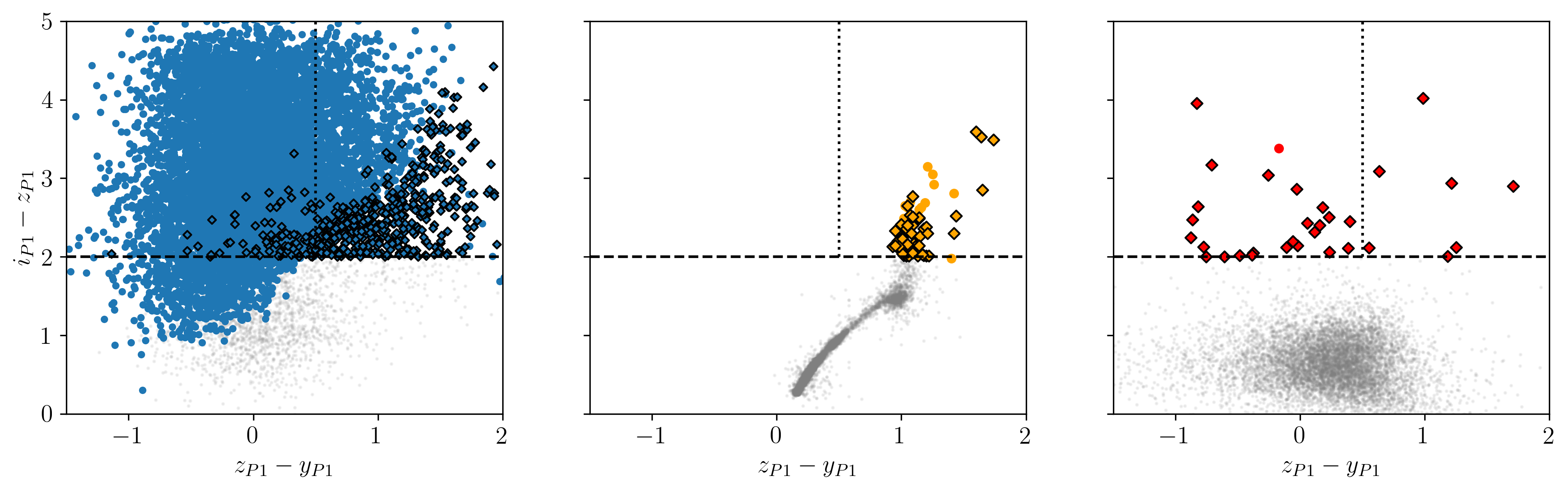}
    \caption{Distribution of sources in the $z_{\rm{P1}} - y_{\rm{P1}}$ vs $i_{\rm{P1}} - z_{\rm{P1}}$ diagram of the test samples of HzQs (left), brown dwarfs (middle), and galaxies (right). Sources not selected with any method are marked in grey. Sources selected with $i-z > 2$ colour cut (dashed line) but rejected by the probability selection are marked as a diamond shape with a black edge. Notably, the $z-y < 0.5$ colour cut (dotted line) rejects all brown dwarfs from the test sample, but also a large portion of the simulated quasar sample.}
    \label{fig:sim_prob}
\end{figure*}

In Figure \ref{fig:sim_prob} the results of the colour and probability selection on the test samples of HzQs (left) brown dwarfs (middle) and galaxies (right) are summarised. Here it is clear that the probability selection method handily rejects contaminants that occupy the same colour space as HzQs and would normally be included in colour selection. Notable is that though both methods recover a similar amount of HzQs, different subsets are selected, as the probability selection recovers a significant fraction of HzQs below the colour cut, while also rejecting a portion of HzQs above the colour cut that lies close to the colour space of brown dwarfs. As HzQs with  $i-z < 2$ have generally lower redshifts, this demonstrates that the probability selection can be especially effective in recovering HzQs at $z\sim 5.6$.

\section{Spectroscopic follow-up}
\label{sec:spec_obs}
To demonstrate the efficacy of our quasar selection method and confirm the nature of the candidate sources, we obtained spectroscopy primarily targeting the Ly$\alpha$ lines for the most promising HzQs identified by our selection method. Our final sample of high quality HzQ candidates, however, still contained an impractical number of sources for additional spectroscopic observations, and therefore we assigned priorities to sources in our final sample based on available independent ancillary data primarily at infrared wavelengths, which were crucially not used in the probability assignment using our method.

We first cross-matched our candidate HzQ sample with the AllWISE Data Release catalogue\footnote{\url{https://wise2.ipac.caltech.edu/docs/release/allwise/}}, which builds upon the data collected by the \textit{Wide-field Infrared Survey Explorer} (\textit{WISE}; \citealt{wright2010}) mission by additionally including data from the NEOWISE surveys \citep{mainzer2011}. The AllWISE data contains photometry in the \textit{WISE} $W1$, $W2$, $W3$ and $W4$ bands, offering wavelength coverage in the range $3.4 - 22\,\mu m$. We additionally cross match our sample with the UKIRT Hemisphere Survey \citep[UHS,][]{dye2018} data release containing deep $J$-band imaging and source catalogues over $\sim12700$ deg$^2$ of the sky.

We used this ancillary data to assign priorities to sources in our candidate HzQ sample for spectroscopic follow-up. First, we assigned higher priority to sources with lower $|z_{P1}-y_{P1}|$ and $|y_{P1}-J|$, which brings us closer to the locus of colours from known HzQs at $z>5.5$. Next, based on the infrared colours seen in the sample of known HzQs from \citet{banados2016}, which was also used as a validation sample for this study, we assigned higher priorities to sources that satisfied the following conditions:
\begin{align*}
    W2 - W3 &> 0 \\
    -0.2 < W1 - W2 &< 0.85 \\
    -0.7 < y_{\rm{P1}} - W1 &< 2.2
\end{align*}{}
Lastly, the brightest sources in our sample were assigned higher priority, purely to make the process of spectroscopic follow-up more efficient. Having assigned a priority for spectroscopic follow up to each candidate HzQ, we now describe our spectroscopic observations below. 

\subsection{Description of observations}

The spectroscopic observations of our candidate HzQs presented in this work were obtained using the Intermediate Dispersion Spectrograph\footnote{\url{http://www.ing.iac.es/astronomy/instruments/ids/}} (IDS) on the 2.5m Isaac Newton Telescope (PI: Wagenveld, Program: N17). The observations were taken over a period of 6 nights in Spring 2019, during which 13 of the highest priority candidate HzQs were observed. Three nights were unfortunately lost due to bad weather, and the remaining three nights had favourable conditions with an average seeing of $0.5"$ from 6 to 8 April 2019.

The observations were taken using the R400R grating in the Red arm of the spectrograph, with a slit width of 1.5 arcseconds and slit length of 3 arcminutes. Standard afternoon calibrations were performed with both lamp and sky flats taken before each observing night. A flux standard was observed at the beginning and the end of each night. We used CuAr+CuNe lamps for wavelength  calibration, which were observed at the position of each target before the sky exposure.

The targets were observed in blocks of 1800s, with total integration times per source ranging from 3600s to 7200s. Due to telescope limitations and higher priority assigned to brighter sources, only sources brighter than a $z$-magnitude of $20.5$ were observed. Blind offsetting was used to acquire faint targets and standard data reduction procedures that include bias subtraction, flat-fielding, sky subtraction, wavelength calibration and flux calibration were performed using a custom \textsc{python} based data reduction pipeline written by our team\footnote{\url{https://github.com/aayush3009/INT-IDS-DataReduction}}, which is based on \textsc{ccdproc} \citep{craig2021}. 

Of the 13 targets observed in this run, 11 could not be conclusively classified based on the spectra obtained. In most cases only very faint continuum was spotted with potential narrow lines. Unfortunately the signal to noise (S/N) of the continuum or the emission lines for these sources was not sufficient to unambiguously determine redshifts or classify the sources as either dwarf stars in the Milky Way or low redshift galaxies. Three of the observed sources had an accompanying radio detection, but did not contain strong emission line features in their spectra. However, the clear absence of a strong Ly$\alpha$ line or a break blueward of Ly$\alpha$ in their spectra indicated that these sources were unlikely to be quasars at $z \gtrsim 5.5$. 

High S/N spectra, however, were obtained for two sources in our sample, one of which was conclusively classified as a brown dwarf star owing to clear, broad absorption features in the continuum arising from molecules such as TiO \citep[e.g.][]{reiners2007}, which often mimic the Ly$\alpha$ break found in the spectra of HzQs.

The other source, PSO J144128.715+502239.463 (shortened to P144+50 hereafter) was convincingly classified as a previously undiscovered, luminous quasar at a redshift of $z=5.66$, and in the following section we describe the observed properties of this newly discovered HzQ.

\subsection{P144+50 -- a luminous quasar at $z\approx5.7$}

\begin{table}[t]
	\centering
	\caption{Observed optical and infrared magnitudes of P144+50 in all filters relevant to probability calculation and priority assignment.}
	\begin{tabular}{l c}
	\hline
	Filter & Magnitude (AB) \\
	\hline \hline
    PS1 $g$ & $>23.2$ \\
   PS1 $r$ & $>23.2$ \\
   PS1 $i$ & $20.71\pm0.03$ \\
   PS1 $z$ & $19.31\pm0.02$ \\ 
   PS1 $y$ & $19.41\pm0.03$ \\
   UHS $J$ & $19.34\pm0.06$ \\
   \textit{WISE} $W1$ & $18.22\pm0.04$ \\
   \textit{WISE} $W2$ & $18.21\pm0.05$ \\
   \textit{WISE} $W3$ & $17.46\pm0.28$ \\
    \hline
    \end{tabular}
    \label{table:p144_mag}
\end{table}
\begin{figure*}
\resizebox{\hsize}{!}
        {\includegraphics{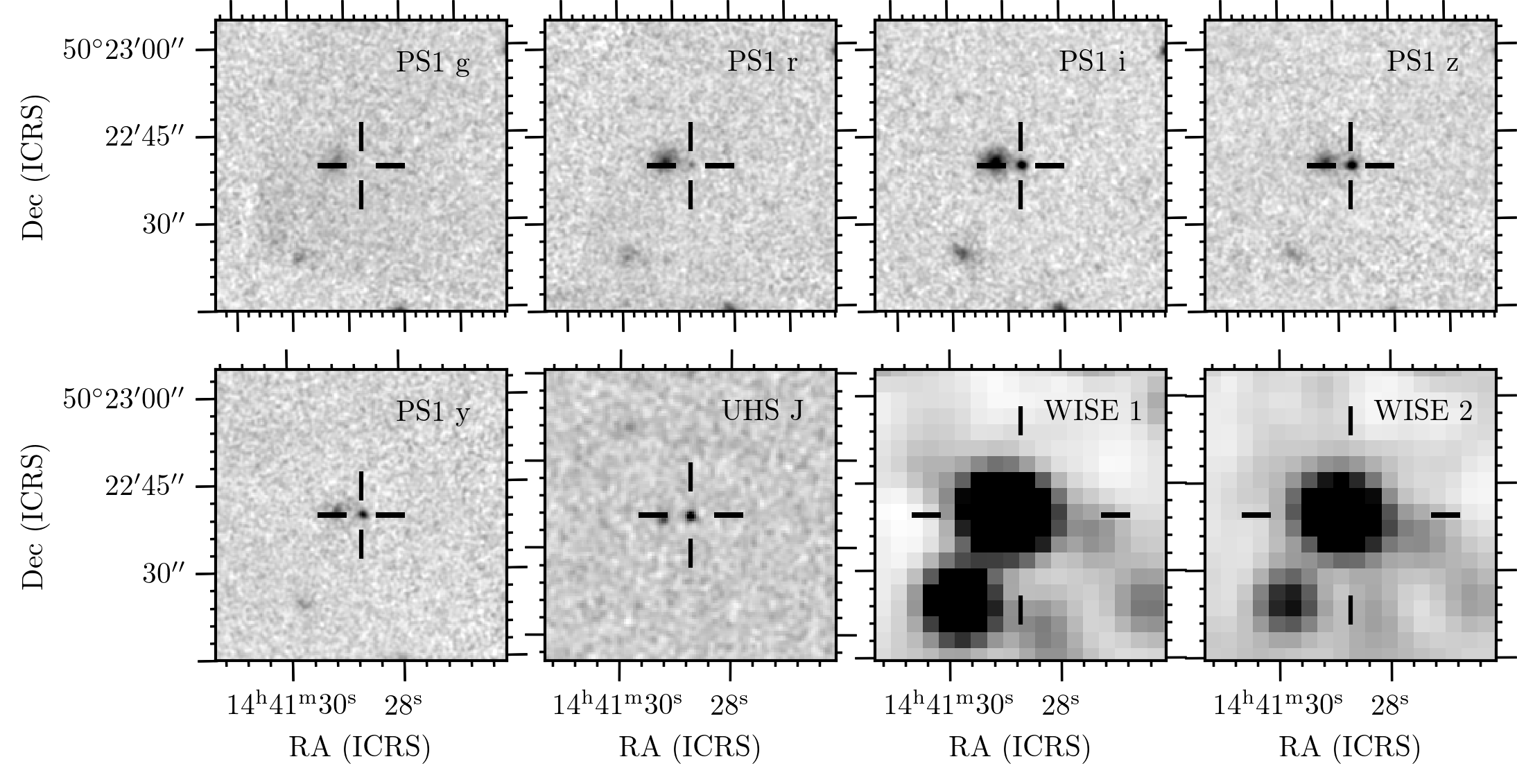}}
  \caption{P144+50 in optical and infrared bands. In the WISE images it is blended with the neighbouring galaxy. Given the fact that the quasar is brighter in those bands while the galaxy drops off, and that its WISE magnitudes are consistent with that of other HzQs, it is likely that most of the emission in the WISE bands comes from the quasar.}
     \label{fig:quasarbands}
\end{figure*}

The most luminous and promising source of the candidate sample, P144+50 has a very clear point-source like structure across the available broad band photometry, as illustrated in Figure \ref{fig:quasarbands}. This luminous quasar was most likely missed by earlier searches owing to its relatively low $i_{P1}-z_{P1}$ colour of $1.4$, which may be excluded by traditional binary colour cuts. Although no radio counterpart for this quasar was identified within the LoTSS DR1 catalogues, P144+50 was still assigned a probability of $P_q = 0.01$ from our method, demonstrating that our novel HzQ selection method is capable of assigning realistically high probabilities even to non-radio detected HzQs. In Table \ref{table:p144_mag} we give the apparent magnitudes of P144+50 in the available optical and infrared filters. 

\begin{figure*}
\resizebox{\hsize}{!}
        {\includegraphics{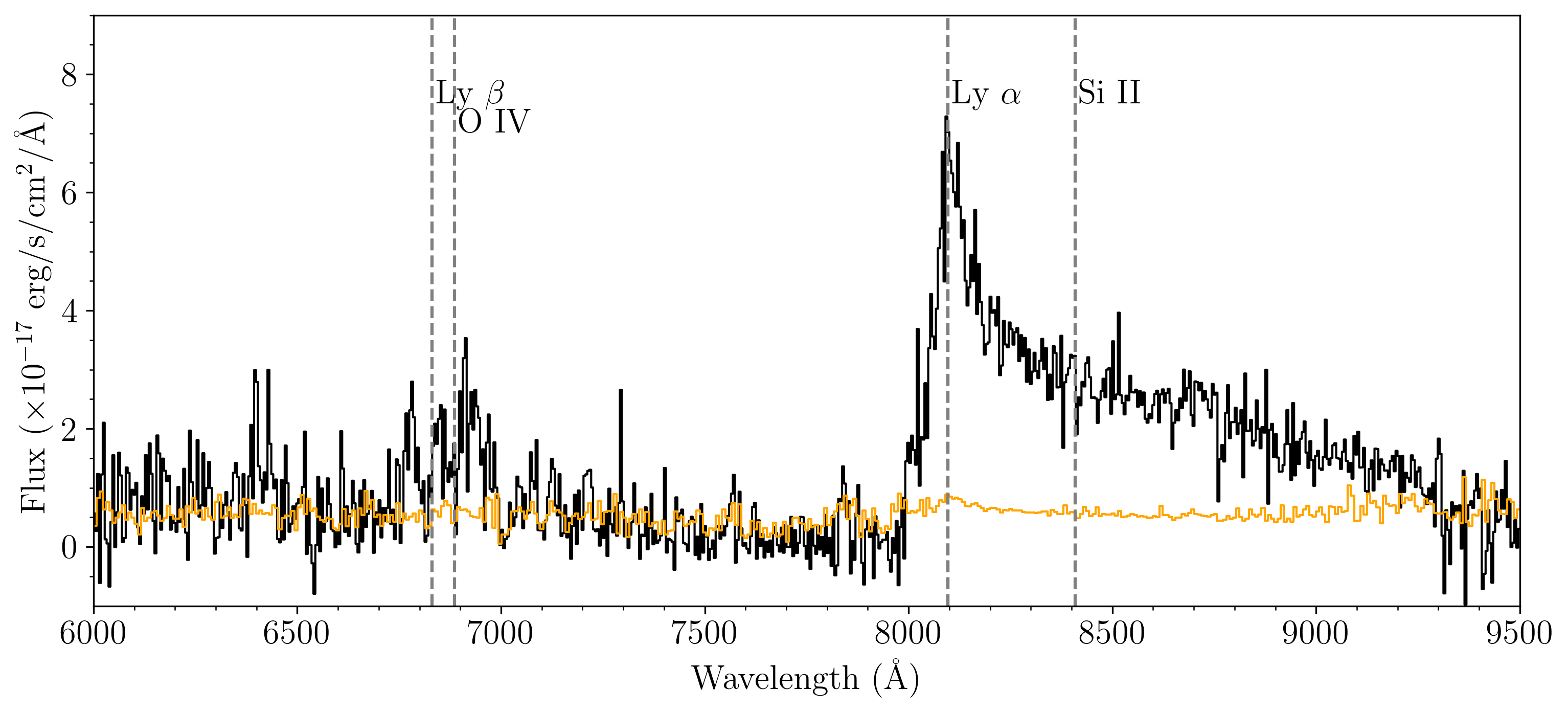}}
  \caption{1D spectrum of P144+50 taken using the IDS instrument on the 2.5m Isaac Newton Telescope (INT). The characteristic break blueward of the Ly$\alpha$ line is clearly visible, unambiguously identifying it as a high redshift quasar with a redshift of $z=5.66$ based on its Ly$\alpha$ emission. Some flux is seen around the expected Ly$\beta$ and O\,\textsc{IV} features. Besides Ly$\alpha$, there are no clear emission lines visible in the spectrum due to the limited sensitivity of the INT. The orange line indicates the noise level of the spectrum.}
     \label{fig:quasarspec}
\end{figure*}

The 1D spectrum of P144+50, shown in Figure \ref{fig:quasarspec} displays a bright and broad strong Ly$\alpha$ feature, with a clear break in the spectrum blueward of the line, showing the Gunn-Peterson trough. The peak of the Ly$\alpha$ line suggests a redshift of $z=5.66$. No other rest-frame UV emission features are identified, however the Si\,\textsc{ii} absorption feature may be present. 

We additionally detect a Ly$\alpha$ forest, and faint signs of the presence of an ionised proximity zone around this QSO. Additional flux is detected around the rest-frame Ly$\beta$ and O \textsc{iv} wavelengths. Unfortunately, due to the limited S/N of our INT observations any meaningful constraints on either the proximity zone or the neutrality of the intervening IGM along the line-of-sight due from the Ly$\alpha$ forest cannot be derived. Therefore, deeper follow-up observations with larger telescopes facilities are required to draw robust conclusions.

Demonstrating perfectly the merits of the new HzQ selection method introduced in this paper, P144+50 is a bright, hitherto undiscovered quasar at $z=5.66$. Its  rest-frame UV magnitude, $M_{1450} = -27.22$, puts it at the brighter end of the quasar luminosity function at $z\sim6$ \citep{manti2017} and amongst the most luminous quasars currently known at this redshift. Given that the sky area covered by our search is ${\sim}1\%$ of the full sky, and the sensitivity of spectroscopic observations restringing the follow up to only those sources with magnitudes brighter than 20.5\,AB, it is not improbable that other such undiscovered quasars exist within the large sky surveys, which may have been missed searches relying on binary colour selection. 

Additionally, the P144+50 lies within the redshift range investigated by \citet{yang2017}, demonstrating that our GMM-based HzQ selection approach is able to successfully identify quasars within the so-called `redshift gap' often encountered by HzQ searches employing optical broadband selection. Thanks to the increased discriminatory power provided by our algorithm, GMM-based HzQ searches might provide a powerful method for more complete samples of HzQs, including those that lie within the redshift gaps in ground-based optical broadband searches.

\section{Future prospects}
\label{sec:discussion}

The Bayesian quasar selection method presented in this work is built from priors informed by empirical relations and likelihood models from machine learning, as opposed to binary cuts in optical/infrared magnitudes and colours. As a result, this method relies heavily on the accuracy of priors derived for both HzQ populations as well as common contaminants in HzQ searches. Therefore, the priors can be improved in an iterative fashion by folding in the results from the ever increasing spectroscopic confirmation of candidate HzQs selected from photometric surveys. 

The inclusion of additional photometric data can also help to improve the priors, resulting in a more accurate HzQ probability assignment. For example, $J$ band and \textit{WISE} photometry for known HzQs and contaminants can be used to improve the estimation of priors, which for this work have only been used to shortlist candidate HzQs for spectroscopic follow up. Expanding the model to include these additional dimensions can further increase its precision and reliability due to the increased colour information available, as well as extend its application for searches of HzQ at even higher redshifts. 

The range of redshifts selected for this analysis ($5.6 < z < 6.5$) was selected to enable validation with known HzQs that were selected using $i-z$ photometric colours. However, given that samples of HzQs can be simulated for training and existing HzQs can be used as validation for our selection method, other redshift ranges can be probed. For example, searches for HzQs at $z> 6.5$ can be readily carried out using our algorithm by using a handful of known HzQs at these redshifts, selected based on their $z-y$ colours. Although the efficacy of the method may not be as high for selecting $z>6.5$ HzQ candidates, owing to the lower number of HzQs known at these redshifts that could be used for training and validation. As mentioned earlier, the inclusion of more photometric data may help improve the priors for HzQs at the highest redshifts. 

As the method should be extendable to larger datasets, one potential bottleneck will be excising the remaining unwanted sources after assigning probability. For the sample described in this paper this final step was done through visual inspection of 263 high probability candidates. For much larger initial sample size, this method is no longer feasible. From the visual inspection we performed, most of these unwanted sources were rejected on grounds of either being spurious, having incorrect magnitude, or appearing extended. Spurious sources are essentially removed if we force all sources to have a counterpart in WISE and/or UHS. As these magnitudes are used anyway assign priority to sources (Section \ref{sec:spec_obs}), it will be doubly advantageous to implement such a selection. Incorrect magnitudes can be remedied by performing photometry directly on the Pan-STARRS images. Finally, there is clear need to differentiate between point sources and extended sources. There are several ways to do this with the Pan-STARRS catalogues, such as comparing PSF and Kron magnitudes of sources\footnote{\url{https://outerspace.stsci.edu/display/PANSTARRS/How+to+separate+stars+and+galaxies}} \citep{farrow2013}. These methods of differentiating between point sources and extended sources however become less reliable towards lower magnitudes, where we expect more quasars. We note here that efforts are currently underway to use machine learning techniques to morphologically classify radio sources \citep[e.g.][]{Mostert21}, which could suitably be extended to morphological classification of candidate HzQs from optical images. 

We have shown the discriminatory power of our HzQ selection method and demonstrated that it is possible to shortlist manageable numbers of high quality HzQ candidates from large photometric data sets. With the aforementioned flexibility and room for improvement, our algorithm can potentially be applied to even larger, deeper surveys of the sky enabled by existing state-of-the-art and upcoming ground- and space-based optical and infrared observatories such as the Vera C. Rubin Observatory \citep{ivezic2019}, \textit{Euclid} \citep{laureijs2011}, the \textit{Nancy Grace Roman Space Telescope} \citep[formerly known as \textit{WFIRST};][]{spergel2015}, and existing large surveys such as the Kilo-Degree Survey \citep[KiDS;][]{dejong2013} and Dark Energy Survey \citep[DES;][]{thedarkenergysurveycollaboration2005} to name a few. 

Finally, while the $z=5.66$ quasar discovered in this analysis is undetected in LoTSS radio continuum imaging, the high detection fraction of known $z > 5$ sources within the 5700 deg$^2$ of the forthcoming LoTSS Data Release 2 \citep[36\% at $>2\sigma$ significance;][]{gloudemans2021} illustrates that the radio continuum observations can provide valuable additional information for HzQ selection and remains a powerful tool to crucially exclude contamination from Galactic dwarf stars. Relatively shallow but large area existing radio surveys such as FIRST \citep{vla_first} and NVSS \citep{vla_nvss} carried out with the Very Large Array (VLA) that have led to the discovery of several radio-loud quasars at $z\gtrsim5$ \citep[e.g.][]{banados2015}, and TGSS Alternative Data Release \citep{Intema2017} covering $\sim 37000$ sq/ deg. of the sky at 150 MHz, which has already led to the discovery of the most distant radio selected galaxy currently known \citep{saxena2018}. The full LoTSS data release will offer sensitive radio coverage over very large sky areas over the northern hemisphere, enabling the inclusion of radio priors for a large number of candidate HzQs. These sky areas and sensitivities will be improved by upcoming ultra-deep radio surveys such as those by the Square Kilometre Array \citep[SKA; ][]{Dewdney2009} and its precursors like MeerKAT \citep{jonas2016} and ASKAP \citep{hotan2021} enabling even fainter radio detections. 

Therefore, the HzQ selection method presented in this work is flexible, and has room for improvement given the availability of deep photometric data over large parts of the sky via existing and future large area sky surveys across wavelengths. Our method presents also provides a robust framework within which the additional radio information can be incorporated to potentially identify even radio-faint quasars in the early Universe.

\section{Summary and conclusions}
\label{sec:conclusions}

In this paper we have presented a novel method for selecting candidate high redshift quasars (HzQs; $z\gtrsim5$) from large photometric data sets, making use of informed priors and Gaussian mixture models (GMMs) within a Bayesian framework. Our method attempts to capture the HzQ population more completely compared to traditionally used binary cuts in optical magnitudes and colours, while minimising the likelihood of contamination from foreground sources such as dwarf stars in the Milky Way and lower redshift dusty galaxies. 

Our novel selection method builds upon previous works employing Bayesian selection of HzQ candidates using informed priors. The novelty of our methods lies in using GMMs to obtain likelihoods in optical colour-colour spaces using photometry of populations of known and simulated HzQs, as well as common contaminants such as M, L and T brown dwarf stars and low redshift dusty galaxies that often mimic the observed optical photometric properties of HzQs. Additional priors based on the security of optical detections, respective sky densities of the source populations as well as a radio detection are used to calculate the probability of a particular source detected in a large photometric sky survey being a candidate HzQ.

We run our GMM based HzQ search method on photometric data from the publicly available Pan-STARRS DR1 (PS1) over a limited area on the sky, coinciding with deep radio imaging from LoTSS in the HETDEX Spring field covering $\sim400$ square degrees. Using in particular photometry in the PS1 $i$, $z$ and $y$ bands, we assign candidate HzQ probabilities to $\sim5\times10^5$ sources from PS1. Adopting a HzQ posterior probability threshold that results in the selection of $\sim90\%$ of known HzQs at $z\gtrsim5.5$ and the rejection of $\gtrsim 99\%$ of known foreground contaminants such as dwarf stars or low redshift galaxies, we shortlist 263 candidate HzQs with high probabilities. By visually inspecting these candidates to spot any obvious artefacts, we select 63 sources in the final high probability candidate HzQ sample, which can subsequently be followed up spectroscopically. To test the efficacy of the method, we run the probability selection on test samples of simulated HzQs and previously used samples of dwarfs an galaxies. We find that the efficacy of the probability method is higher than traditional colour cuts, decreasing the fraction of accepted contaminants by 86\% while retaining a similar fraction of HzQs. While more stringent colour cuts decrease the contaminant fraction to levels similar to that of the probability selection, less HzQs are recovered. The efficacy of the probability selection is increased further once radio data is taken into account, reducing the fraction of contaminants by 99\% compared to the traditional colour cut at the cost of selecting only quasars that have a radio detection.

Follow-up spectroscopic observations were then carried out for the highest priority HzQ candidates from our sample, with 13 candidates targeted with the 2.5m Isaac Newton Telescope. Although the nature of 11 out of these 13 candidates could not be confirmed owing to low signal-to-noise ratios in the relatively shallow spectra, a lack of strong Ly$\alpha$ emission or Ly$\alpha$ absorption present in the spectrum was used to rule out a very high redshift nature. 

However, the exact nature of 2 candidates could be established, with one being a brown dwarf star and the other being a previously undiscovered, luminous quasar at $z=5.66$ (P144+50). The spectrum of P144+50 shows a strong and broad Ly$\alpha$ line, with a strong break in the spectrum bluewards of Ly$\alpha$ indicative of a Gunn-Peterson trough. P144+50 has a rest-frame UV magnitude of $M_{1450}=-27.22$, putting it at the very bright end of the luminosity function at this redshift. This HzQ was likely missed by earlier searches owing to its $i-z$ photometric colour of $1.4$, which falls below the traditional limits requiring $i-z > 1.5$.

The discovery of this previously undiscovered, luminous quasar at $z=5.66$ serves as a validation of our novel HzQ selection method, indicating that a probabilistic method of selecting HzQs from large photometric surveys may perform better at returning more complete samples of HzQs as opposed to binary selections based on cuts in optical/infrared colours or magnitudes. Our method has room for improvement with the inclusion of more photometric data when calculating posterior probabilities, and as such can be employed on larger incoming sky surveys to discover new quasars, into the epoch of reionisation.

\begin{acknowledgements}
This paper is dedicated to the memory of our dear friend and collaborator Maolin Zhang whose contributions to this project, and the wider field of astronomy, were tragically cut short.

KJD acknowledges funding from the European Union’s Horizon 2020 research and innovation programme under the Marie Sk\l{}odowska-Curie grant agreement No. 892117 (HIZRAD).

We thank the anonymous referee for their useful comments and feedback.

The Pan-STARRS1 Surveys (PS1) and the PS1 public science archive have been made possible through contributions by the Institute for Astronomy, the University of Hawaii, the Pan-STARRS Project Office, the Max-Planck Society and its participating institutes, the Max Planck Institute for Astronomy, Heidelberg and the Max Planck Institute for Extraterrestrial Physics, Garching, The Johns Hopkins University, Durham University, the University of Edinburgh, the Queen's University Belfast, the Harvard-Smithsonian Center for Astrophysics, the Las Cumbres Observatory Global Telescope Network Incorporated, the National Central University of Taiwan, the Space Telescope Science Institute, the National Aeronautics and Space Administration under Grant No. NNX08AR22G issued through the Planetary Science Division of the NASA Science Mission Directorate, the National Science Foundation Grant No. AST–1238877, the University of Maryland, Eotvos Lorand University (ELTE), the Los Alamos National Laboratory, and the Gordon and Betty Moore Foundation. 

This publication makes use of data products from the Wide-field Infrared Survey Explorer, which is a joint project of the University of California, Los Angeles, and the Jet Propulsion Laboratory/California Institute of Technology, and NEOWISE, which is a project of the Jet Propulsion Laboratory/California Institute of Technology. WISE and NEOWISE are funded by the National Aeronautics and Space Administration.

LOFAR data products were provided by the LOFAR Surveys Key Science project (LSKSP; \url{https://lofar-surveys.org/}) and were derived from observations with the International LOFAR Telescope (ILT). LOFAR \citep{lofar} is the Low Frequency Array designed and constructed by ASTRON. It has observing, data processing, and data storage facilities in several countries, which are owned by various parties (each with their own funding sources), and which are collectively operated by the ILT foundation under a joint scientific policy. The efforts of the LSKSP have benefited from funding from the European Research Council, NOVA, NWO, CNRS-INSU, the SURF Co-operative, the UK Science and Technology Funding Council and the J\"{u}lich Supercomputing Centre.

The Isaac Newton Telescope is operated on the island of La Palma by the Isaac Newton Group of Telescopes in the Spanish Observatorio del Roque de los Muchachos of the In-stituto de Astrof\'{i}sica de Canarias.

This research has made extensive use of TOPCAT \citep{topcat}.  

The algorithm presented in this work is written in \textsc{python} and will be made publicly available. In the mean time the code will be shared upon reasonable written requests to the authors.

\end{acknowledgements}

\bibliographystyle{aa}
\bibliography{HzQ-paper}

\newpage
\begin{appendix}
\onecolumn
\section{HzQ candidates}
\begin{longtable}{lllllll}
\caption{High probability HzQ candidates}
\label{tab:candidates}
\\
\hline
Name   & $i_{P1}$   & $z_{P1}$    & $y_{P1}$     & LoTSS Flux   & $P_q$ & Observed  \\
       &            &             &              & (mJy)        &    &           \\
\hline \hline
\endfirsthead
\caption{continued}\\
\hline
Name   & $i_{P1}$   & $z_{P1}$    & $y_{P1}$     & LoTSS Flux   & $P_q$  & Observed \\
       &            &             &              & (mJy)        &     &          \\
\hline \hline
\endhead
\endfoot
\hline
\endlastfoot
PSO J151528.1+421313.8 & $21.49\pm0.11$     & $19.58\pm0.04$        & $20.03\pm0.18$    &                        & $1.6\times10^{-3}$ &  \\
PSO J115421.7+421840.7 & $21.68\pm0.12$     & $19.32\pm0.06$        & $20.64\pm0.17$    &                        & $5.8\times10^{-1}$ &  \\
PSO J150748.8+422307.8 & $21.50\pm0.06$     & $19.45\pm0.00$        & $19.90\pm0.26$    &                        & $3.0\times10^{-3}$ &  \\
PSO J123718.4+422839.6 & $22.27\pm0.04$     & $20.09\pm0.18$        & $20.63\pm0.16$    &                        & $1.4\times10^{-3}$ &  \\
PSO J124208.8+423946.4 & $22.15\pm0.01$     & $20.04\pm0.15$        & $20.21\pm0.13$    &                        & $8.4\times10^{-4}$ & Yes \\
PSO J124911.0+425105.3 & $22.11\pm0.05$     & $20.13\pm0.15$        & $20.98\pm0.21$    &                        & $8.3\times10^{-4}$ &  \\
PSO J125047.9+430833.7 & $22.17\pm0.01$     & $19.55\pm0.09$        & $19.66\pm0.14$    &                        & $1.3\times10^{-1}$ &  \\
PSO J123203.0+432745.0 & $21.94\pm0.14$     & $19.64\pm0.17$        & $20.47\pm0.15$    &                        & $4.4\times10^{-3}$ &  \\
PSO J124221.7+434033.2 & $21.70\pm0.06$     & $20.27\pm0.00$        & $20.65\pm0.11$    &                        & $1.4\times10^{-3}$ &  \\
PSO J122900.5+441359.8 & $21.61\pm0.05$     & $19.39\pm0.16$        & $20.24\pm0.17$    &                        & $8.8\times10^{-3}$ &  \\
PSO J121800.4+453150.9 & $21.78\pm0.17$     & $19.38\pm0.10$        & $20.53\pm0.15$    &                        & $5.2\times10^{-2}$ &  \\
PSO J120837.0+454149.4 & $22.15\pm0.17$     & $19.79\pm0.16$        & $20.55\pm0.17$    &                        & $2.9\times10^{-3}$ &  \\
PSO J114519.9+454428.0 & $21.63\pm0.04$     & $19.93\pm0.00$        & $20.80\pm0.17$    &                        & $2.1\times10^{-2}$ &  \\
PSO J112111.5+461150.9 & $21.47\pm0.17$     & $19.55\pm0.11$        & $20.87\pm0.19$    &                        & $1.5\times10^{-3}$ &  \\
PSO J142738.5+473727.4 & $21.24\pm0.11$     & $20.70\pm0.02$        & $21.38\pm0.00$    &                        & $1.1\times10^{-2}$ &  \\
PSO J150321.1+480022.9 & $21.84\pm0.07$     & $17.61\pm0.01$        & $20.50\pm0.19$    &                        & 1.0               &  \\
PSO J151021.5+490023.1 & $21.69\pm0.27$     & $18.21\pm0.02$        & $20.70\pm0.18$    &                        & 1.0               &  \\
PSO J144128.7+502239.4 & $20.71\pm0.03$     & $19.31\pm0.02$        & $19.41\pm0.03$    &                        & $1.1\times10^{-2}$ & Yes \\
PSO J112418.7+504151.3 & $21.78\pm0.05$     & $20.08\pm0.00$        & $20.73\pm0.18$    &                        & $4.7\times10^{-3}$ &  \\
PSO J152639.5+520303.0 & $21.44\pm0.02$     & $17.29\pm0.01$        & $19.90\pm0.16$    &                        & 1.0              &  \\
PSO J144047.0+520934.6 & $21.00\pm0.10$     & $18.95\pm0.04$        & $20.44\pm0.17$    &                        & $8.3\times10^{-1}$ &  \\
PSO J121906.9+524229.8 & $21.35\pm0.03$     & $19.90\pm0.02$        & $19.81\pm0.06$    &                        & $5.2\times10^{-4}$ & Yes \\
PSO J120853.9+540651.1 & $21.37\pm0.05$     & $19.92\pm0.00$        & $20.53\pm0.19$    &                        & $8.2\times10^{-4}$ &  \\
PSO J110945.2+574348.4 & $21.62\pm0.04$     & $17.88\pm0.01$        & $19.73\pm0.17$    &                        & 1.0              &  \\
PSO J112328.2+595614.9 & $21.42\pm0.08$     & $18.70\pm0.00$        & $20.52\pm0.18$    &                        & 1.0              &  \\
PSO J135335.3+600430.6 & $21.51\pm0.04$     & $20.73\pm0.04$        & $21.69\pm0.00$    &                        & $2.5\times10^{-3}$ &  \\
PSO J152721.9+610352.3 & $21.08\pm0.14$     & $18.79\pm0.01$        & $18.19\pm0.01$    &                        & $7.3\times10^{-3}$ & Yes \\
PSO J141715.5+615224.3 & $22.00\pm0.16$     & $19.52\pm0.15$        & $20.54\pm0.17$    &                        & $1.7\times10^{-2}$ &  \\
PSO J112052.2+472605.0 & $21.36\pm0.07$     & $19.99\pm0.12$        & $20.71\pm0.16$    & $1.10  \pm 0.10 $      & $3.4\times10^{-3}$ &  \\
PSO J141837.2+474852.2 & $22.65\pm0.35$     & $21.32\pm0.11$        & $20.43\pm0.19$    & $3.29  \pm 0.11 $      & $7.2\times10^{-4}$ &  \\
PSO J113104.0+475003.9 & $21.64\pm0.17$     & $19.67\pm0.12$        & $19.86\pm0.15$    & $0.44  \pm 0.10 $      & $2.8\times10^{-2}$ & Yes \\
PSO J123823.6+475933.1 & $21.13\pm0.05$     & $19.91\pm0.18$        & $20.41\pm0.15$    & $0.69  \pm 0.15 $      & $2.1\times10^{-3}$ &  \\
PSO J131244.6+495724.5 & $21.62\pm0.22$     & $20.28\pm0.16$        & $20.23\pm0.19$    & $0.43  \pm 0.09 $      & $7.8\times10^{-4}$ & Yes \\
PSO J123626.6+501036.9 & $21.51\pm0.08$     & $20.16\pm0.20$        & $20.97\pm0.41$    & $0.64  \pm 0.09 $      & $2.3\times10^{-3}$ &  \\
PSO J124654.9+501623.7 & $21.71\pm0.19$     & $20.65\pm0.13$        & $19.35\pm0.00$    & $0.67  \pm 0.11 $      & $1.2\times10^{-3}$ &  \\
PSO J112037.6+502404.9 & $21.82\pm0.20$     & $20.81\pm0.16$        & $19.44\pm0.22$    & $0.45  \pm 0.12 $      & $6.7\times10^{-4}$ &  \\
PSO J134157.7+512952.2 & $21.96\pm0.19$     & $20.93\pm0.12$        & $19.49\pm0.16$    & $31.89 \pm 0.16 $      & $7.8\times10^{-4}$ &  \\
PSO J130926.4+525922.1 & $21.82\pm0.20$     & $20.70\pm0.17$        & $19.50\pm0.17$    & $0.97  \pm 0.10 $      & $8.0\times10^{-4}$ &  \\
PSO J113311.2+420443.2 & $21.50\pm0.04$     & $19.68\pm0.00$        & $20.17\pm0.11$    &                        & $1.4\times10^{-3}$ &  \\
PSO J123740.1+420851.0 & $22.06\pm0.14$     & $20.13\pm0.19$        & $20.63\pm0.16$    &                        & $7.3\times10^{-4}$ &  \\
PSO J151948.4+423446.7 & $21.99\pm0.17$     & $19.96\pm0.00$        & $19.88\pm0.18$    &                        & $2.1\times10^{-3}$ & Yes \\
PSO J124059.8+431019.5 & $22.14\pm0.18$     & $20.26\pm0.17$        & $20.74\pm0.17$    &                        & $7.0\times10^{-4}$ &  \\
PSO J140022.9+433822.2 & $22.11\pm0.30$     & $19.56\pm0.02$        & $18.40\pm0.02$    &                        & $1.2\times10^{-3}$ &  \\
PSO J145612.6+442417.2 & $21.71\pm0.01$     & $20.12\pm0.20$        & $20.12\pm0.10$    &                        & $5.4\times10^{-4}$ & Yes \\
PSO J114416.9+443451.0 & $21.36\pm0.05$     & $19.50\pm0.15$        & $19.70\pm0.15$    &                        & $2.5\times10^{-3}$ & Yes \\
PSO J135622.5+453320.5 & $21.83\pm0.18$     & $19.88\pm0.00$        & $19.99\pm0.18$    &                        & $1.5\times10^{-3}$ &  \\
PSO J123757.3+465507.2 & $21.73\pm0.20$     & $20.25\pm0.04$        & $20.61\pm0.16$    &                        & $5.7\times10^{-4}$ &  \\
PSO J124016.5+473737.9 & $21.58\pm0.03$     & $19.55\pm0.10$        & $20.39\pm0.24$    &                        & $1.7\times10^{-3}$ &  \\
PSO J124300.0+481418.3 & $22.81\pm0.28$     & $20.96\pm0.29$        & $20.47\pm0.17$    &                        & $7.4\times10^{-4}$ &  \\
PSO J124203.1+495354.1 & $21.44\pm0.05$     & $19.91\pm0.20$        & $20.62\pm0.18$    &                        & $6.3\times10^{-4}$ &  \\
PSO J124656.0+503223.3 & $21.47\pm0.04$     & $19.96\pm0.17$        & $20.63\pm0.16$    &                        & $6.2\times10^{-4}$ &  \\
PSO J131523.9+513827.7 & $21.57\pm0.07$     & $18.58\pm0.02$        & $19.99\pm0.14$    &                        & $2.7\times10^{-3}$ &  \\
PSO J143229.1+534741.1 & $21.94\pm0.18$     & $20.18\pm0.00$        & $20.68\pm0.20$    &                        & $7.4\times10^{-4}$ &  \\
PSO J124450.2+585817.8 & $21.37\pm0.09$     & $19.73\pm0.13$        & $20.33\pm0.08$    &                        & $9.4\times10^{-4}$ &  \\
PSO J144448.1+600520.3 & $21.46\pm0.08$     & $20.04\pm0.17$        & $20.46\pm0.20$    &                        & $6.1\times10^{-4}$ &  \\
PSO J131842.1+600706.5 & $21.67\pm0.08$     & $19.67\pm0.00$        & $20.45\pm0.19$    &                        & $1.7\times10^{-3}$ &  \\
PSO J133347.7+603212.6 & $21.62\pm0.08$     & $20.01\pm0.33$        & $19.99\pm0.04$    &                        & $6.1\times10^{-4}$ & Yes \\
PSO J132006.7+605705.4 & $21.92\pm0.18$     & $19.81\pm0.13$        & $20.12\pm0.14$    &                        & $2.6\times10^{-3}$ &  \\
PSO J131356.0+614833.6 & $21.65\pm0.11$     & $20.12\pm0.17$        & $20.33\pm0.17$    &                        & $5.7\times10^{-4}$ &  \\
PSO J120732.9+492944.0 & $21.50\pm0.07$     & $20.63\pm0.42$        & $19.16\pm0.05$    & $1.16 \pm 0.10 $       & $5.6\times10^{-4}$ & Yes \\
PSO J115605.5+444105.3 & $21.68\pm0.05$     & $19.33\pm0.21$        & $20.76\pm0.18$    &                        & $8.2\times10^{-4}$ &  \\
PSO J105545.9+445655.8 & $21.69\pm0.11$     & $19.77\pm0.26$        & $19.71\pm0.06$    &                        & $2.6\times10^{-3}$ & Yes \\
PSO J130519.1+464845.5 & $21.86\pm0.20$     & $20.22\pm0.18$        & $20.59\pm0.17$    &                        & $5.7\times10^{-4}$ &  \\
PSO J124059.4+483522.9 & $21.70\pm0.18$     & $19.71\pm0.23$        & $19.92\pm0.18$    &                        & $3.0\times10^{-3}$ & Yes \\
PSO J150531.3+610408.5 & $21.74\pm0.10$     & $20.02\pm0.13$        & $19.70\pm0.20$    &                        & $7.5\times10^{-4}$ &  \\
\end{longtable}
\end{appendix}

\end{document}